%% file: paper.tex
\documentclass[12pt]{article}
\usepackage{times}
\usepackage[numbers,sort&compress]{natbib}
\usepackage{listings}
\usepackage{hyperref}
\usepackage{amsmath}
\usepackage{amsthm}
\usepackage{amssymb}
\usepackage{semantic}
\usepackage{graphicx}
\usepackage{tabularx}
\usepackage{wasysym}
\usepackage{rotating}

\usepackage[T1]{fontenc}
\usepackage[scaled]{helvet}

\newcommand{\CPPOX}[0]{C{+$\!$+}0x}
\newcommand{\cppox}[0]{C{\footnotesize +$\!$+}0x}
\newcommand{\CPP}[0]{C{+$\!$+}}
\newcommand{\cpp}[0]{C{\footnotesize +$\!$+}}
\newcommand{\code}[1]{\lstinline{#1}}

\input{defs}

\title{The \CPPOX{} ``Concepts'' Effort}
\author{Jeremy G. Siek}

\date{}

\begin{document}

\maketitle

\begin{abstract}
  \cppox{} is the working title for the revision of the ISO standard
  of the \cpp{} programming language that was originally planned for
  release in 2009 but that was delayed to 2011.  The largest language
  extension in \cppox{} was ``concepts'', that is, a collection of
  features for constraining template parameters.
%
%
  In September of 2008, the \cpp{} standards committee voted the
  concepts extension into \cppox{}, but then in July of 2009, the
  committee voted the concepts extension back out of \cppox{}.

  This article is my account of the technical challenges and debates
  within the ``concepts'' effort in the years 2003 to 2009.
  To provide some background, the article also describes the design
  space for constrained parametric polymorphism, or what is
  colloquially know as constrained generics.
  While this article is meant to be generally accessible, the writing
  is aimed toward readers with background in functional programming
  and programming language theory. This article grew out of a lecture
  at the Spring School on Generic and Indexed Programming at the
  University of Oxford, March 2010.

\end{abstract}




\section{Introduction}
\label{sec:intro}

The inclusion of the Standard Template Library
(STL)~\cite{stepa.lee-1994:the.s:TR} into \cpp{} in 1994 transformed
the development of software libraries in \cpp{}.  Before 1994, \cpp{}
library developers primarily applied object-oriented techniques to
facilitate software reuse. Beginning in 1994, library developers began
to use the methodology of \emph{generic programming}, which was
created by Alexander Stepanov, David Musser, and colleagues in the
1980s~\cite{Kapur:1981fx,Musser87,Kershenbaum88,Musser:1988xk}.
Generic programming is a methodology for decoupling algorithms from
the data structures on which they operate, thereby enabling the use of
a single algorithm implementation with many different data structures.
Section~\ref{sec:intro-generic} of this article gives a brief overview
of generic programming and Section~\ref{generic-programming} goes into
depth regarding how generic programming is accomplished in \cpp{}.
Many software methodologies claim to be ``silver bullets''. Generic
programming is not a silver bullet per se; developing generic
libraries is a difficult task and requires considerable training and
mathematical expertise. However, generic programming can provide
significant cost savings in the long run.

Generic programming, as realized in \cpp{}, relies heavily on the the
template language feature to provide type parameterization of
functions and classes. As programmers applied generic programming to
larger and more complex libraries in the late 1990s and early 2000s,
both the developers and users of these libraries began to suffer from
several shortcomings of \cpp{} templates.
\begin{itemize}\addtolength{\itemsep}{-0.5\baselineskip}
\item Minor errors in the use of template libraries produced
  extraordinarily long and complex error messages.
\item Library developers could not easily check that their library
  code adhered to its specification.
\item The run-time behavior of libraries could differ in unexpected
  ways based on usage context because of non-modular rules for name
  resolution.
\item Applications using \cpp{} template libraries suffered long
  compilation times.
\end{itemize}

In 2003, two groups, one at Indiana University and the other at Texas
A\&M University, set out to solve these problems through the addition
of language features for constraining template parameters.  The work
by these two groups ultimately led to a language extension that was
voted into the working draft of the \cpp{} standard on September 2008.
However, in July of 2009, the standards committee voted to remove the
language extension.

The aim of this article is to describe the technical challenges and
debates within the ``concepts'' effort from 2003 to 2009 and to
provide the background for understanding the design trade offs. The
remainder of this introduction starts with an introduction to generic
programming methodology and then gives an overview of the history of
language support for generic programming.

\subsection{Higher-Order Polymorphic Programming}
\label{sec:higher-order-poly}

Generic programming builds on the style of higher-order, polymorphic
programming that was developed in LISP~\cite{McCarthy:1960dz} in the
1960s and that was crystallized in type systems of the 1970s such as
System F of \citet{GIRARD72} and \citet{REYNOLDS74C} and the
Hindley-Milner~\cite{Milner:1978kh} type system of Standard
ML~\cite{Milner:1990fk}.  A canonical example of higher-order,
polymorphic programming is the \code{foldl} function, shown below and
written in ML. The first argument to \code{foldl} is a binary
function, the second argument can be thought of as a running total,
and the third argument is a list. The \code{foldl} function applies
the binary function to each element of the list and the running total.
The two equations below that define \code{foldl} match on the third
argument, with \code{nil} matching the empty list and \code{x::xs}
matching a non-empty list, binding \code{x} to the element at the
front and binding \code{xs} to the rest of the list.
\begin{lstlisting}
fun foldl f y nil = y
| foldl f y (x::xs) = foldl f (f (x,y)) xs
\end{lstlisting}

This \code{foldl} function is quite flexible; it can compute
sums, products, and many other things. In the first line below, we use
addition as the binary function, \code{0.0} as the initial value
for the running total, and \code{[1.0,2.0,3.0,4.0]} as the list. The
result of \code{foldl} in this case is \code{10.0}, which is
the sum of the integers in the list.
\begin{lstlisting}
> foldl op + 0.0 [1.0,2.0,3.0,4.0];
val it = 10.0 : real

> foldl op * 1 [1,2,3,4];
val it = 24 : int

> foldl op @ [] [[1,2],[3],[4,5,6]];
val it = [4,5,6,3,1,2] : int list
\end{lstlisting}
The \code{foldl} function is \emph{polymorphic} with respect to the
element type of the list. In the first example, the elements are real
numbers. In the second, they are integers, and in the third, they are
lists of integers. The \code{foldl} function is \emph{higher-order}
because it takes a function as a parameter, which controls whether
\code{foldl} computes the sum, product, or concatenation in the above
examples.

To be specific, the kind of polymorphism provided in Standard ML and
in System F is called \emph{parametric polymorphism}. This kind of
polymorphism is particularly important because it enables both modular
type checking and separate compilation. A language provides
\emph{modular type checking} when 1) a call to a function, or
similarly, an instantiation of a generic, can be type checked using
only its type and not its implementation and 2) the definition of
function or generic can be type checked in isolation, without using any
information about call sites or points of instantiation.  Modular type
checking is critical for the development of libraries and large
software systems.  Modular type checking shields the client of a
library from the internals of the library. Dually, modular type
checking provides library developers with an automated means to rule
out large classes of bugs before distributing the library, especially
bugs regarding inconsistencies between the library's specification and
implementation.

A language enables \emph{separate compilation} if it is possible to
produce an executable machine program in time proportional to the size
of the source code of the main program only, even though the program
uses many software libraries.  This definition is somewhat indirect
because the more straightforward and traditional definition is
meaningless in the presence of just-in-time compilation. Separate
compilation is critical to the development of large software systems,
as it provides a means to reduce compilation times during the
software development cycle.

System F is a simple model for understanding how to ensure modular
type checking and separate compilation in a language with
parameterized types. The definition of System F is concise enough that
we can discuss the entire language in a few paragraphs. The syntax of
System F is minimal; it only supports two language features: functions
and generics, both of just one parameter.
\[
\begin{array}{llcl}
\text{term variables} & x,y,z \\
\text{type variables} & \alpha,\beta \\
\text{integers} & n \\
\text{types} & \tau & ::= & \mathsf{int} \mid \alpha \mid \tau \to \tau \mid \forall \alpha.\; \tau \\
\text{expressions}&  e &::=& n \mid x \mid \lambda x:\tau .\; e \mid e\;e  \mid \Lambda \alpha.\; e
   \mid e [\tau]
\end{array}
\]
The types of System F includes type variables, function types, and
universal types, which give types to generics. The terms of System F
include term variables, anonymous functions (the $\lambda$ form), function
application (which is the juxtaposition of two expressions, the first
should evaluate to a function and the second, should evaluate to its
argument), anonymous generics (the $\Lambda$ form), and the \emph{explicit
  instantiation} of generics ($e[\tau]$).

The modularity of System F's type system can be seen by inspecting its
definition, shown in Figure~\ref{fig:system-f}. The horizontal bars
should be read as if-then statements. The statements above the bar are
premises and the statement below the bar is the conclusion.  We use
$\Gamma$ to denote an environment, which is a sequence of the in-scope
term variables (with their type bindings) and type variables.  
The type system disallows duplicate type variables in
environments.  We write $x : \tau \in \Gamma$ to mean
that the first binding for $x$ in $\Gamma$ is $\tau$.
More formally, $x : \tau \in \Gamma$ if and only if 
$\Gamma_i = x : \tau$ for some $i$ such that 
there is no $j < i$ where $\Gamma_j = x : \tau'$ for some $\tau'$.
The notation $\Gamma \vdash e : \tau$ means that $e$ is
well typed and has type $\tau$ in the scope defined by $\Gamma$.  The
notation $[\alpha{:=}\tau']\tau$ is for the capture-avoiding
substitution of $\tau'$ for $\alpha$ in $\tau$.

\begin{figure}[tbp]
  \centering
\begin{gather*}
  \inference{}{\Gamma \vdash n : \mathsf{int}} \quad
  \inference{x : \tau \in \Gamma}{\Gamma \vdash x : \tau} \\[2ex]
  \inference{\Gamma,x:\tau  \vdash e : \tau'}{\Gamma  \vdash \lambda x{:}\tau.\, e : \tau \to \tau'} \quad
  \inference{\Gamma  \vdash e_1 : \tau \to \tau' & \Gamma \vdash e_2 : \tau}{\Gamma \vdash e_1 \; e_2 : \tau'} 
   \\[2ex]
  \inference{\Gamma, \alpha \vdash e : \tau & \alpha \notin \Gamma}
     {\Gamma \vdash \Lambda \alpha.\; e : \forall \alpha.\; \tau} \quad
  \inference{\Gamma \vdash e : \forall \alpha.\; \tau}{\Gamma \vdash e[\tau'] : [\alpha{:=}\tau']\tau}
\end{gather*}  
  \caption{The typing rules for System F.}
  \label{fig:system-f}
\end{figure}

To see the modularity of the type system, first look at the rule for
function application, with the conclusion $\Gamma \vdash e_1 \; e_2 :
\tau'$. This rule requires that the parameter type $\tau$ of the
function $e_1$ is the same type as the type of $e_2$. Note that this
rule does not require any knowledge of the body of the function being
applied, just its type $\tau \to \tau'$. Similarly, look at the rule
for instantiating generics, with the conclusion $\Gamma \vdash
e[\tau'] : [\alpha {:=} \tau']\tau$. Again, the rule does not require
any knowledge of the body of the generic. Instead, it just requires
knowledge of the generic's type.

On the flip side, consider the typing rules for creating functions and
generics. A function knows that its input $x$ has type $\tau$, but it
has no information about what value will be bound to $x$.
Analogously, a generic knows that its parameter $\alpha$ is a type, but it
does not know which type will be bound to $\alpha$. The type system
ensures that a well-typed generic does not depend on which type is
bound to $\alpha$, that is, it ensures that the generic will work with
\emph{any} choice for $\alpha$. The way the type system ensures this is
that it considers $\alpha$ different from every other type. (So $\alpha$ is
only equal to itself.)  This property of the type system comes from
its use of syntactic identity to check whether two types are equal.
Consider the rule for function application: the parameter type $\tau$
has to be syntactically identical to the type of the argument $e_2$.
For example, the following term is not well typed (after adding
support for integers to System F).
\[
  \Lambda \alpha.\, \lambda x {:} \alpha{\to}\alpha. \, (x \, 1)
\]
In the application $(x\; 1)$, the parameter type is $\alpha$ but the
argument type is $\mathtt{int}$.
From the preceding discussion, we see that System F provides both
polymorphism and modular type checking: an instantiation of a generic
can be type checked without referring to the generic's body and the
body of a generic can be type checked without referring to any
instantiations.

Furthermore, the design of System F enables separate compilation but
does not require it. The run-time behavior of the body of a generic
may not depend on the instantiating type. Thus, it is possible to emit
machine code that works for all instantiations of the same generic,
modulo some restrictions on the calling conventions.  In particular,
the calling convention for data associated with a type parameter must
be uniform, that is, it cannot depend on the instantiating type.  The
most common uniform representation is ``boxing'', that is, storing
data on the heap and referring to it through a pointer.
The compiler for a programming language with parametric polymorphism
may choose to support separate compilation and use boxing, such as
Java~\cite{Bracha:1998fk} and Objective Caml~\cite{Leroy:2000fu}, or
it may choose not to for the sake of efficiency, such as the
MLton~\cite{Cejtin:2007fk} compiler for Standard
ML~\cite{Milner:1990fk}.
Better yet, a compiler may support separate compilation in general but
sometimes optimize the instantiation of a generic when the body of the
generic is available~\cite{Kennedy:2001uq}.

\subsection{Generic Programming}
\label{sec:intro-generic}

Generic programming is a methodology that builds upon higher-order,
polymorphic programming, scaling it up to deal with large families of
algorithms and data structures. For example, the STL includes a
function named \code{accumulate} that generalizes \code{foldl} to work
on any representation of a sequence, not just linked lists. Going
further, the STL includes a comprehensive set of sorting and searching
functions. Beyond the STL, programmers applied generic programming in
the development of libraries in the areas of computational geometry
\cite{boissonnat99:_cgal}, quantum mechanics~\cite{troyer:_alps},
graph algorithms (the author's Boost Graph
Library~\cite{Siek:2002ly}), and many
more~\cite{Siek:2000ca,koethe99:_reusable_vision,Dawes:qf,pitt01:_bioinf_template_lib}.

The main programming language capability that is needed for
higher-order, polymorphic programming to scale to large software
libraries is the capability to organize and group together parameters,
such as the parameters \code{f} and \code{y} of \code{foldl}. The
number of such parameters grows large for interesting algorithms,
upwards of 20 parameters! The key to organizing these parameters is
that there are relationships between subgroups of parameters, and that
these relationships form coherent abstractions.  For example, in the
use of \code{foldl} with multiplication, the choice of \code{1} as the
argument to parameter \code{y} was not by chance.  The integer
\code{1} is the \emph{identity element} for multiplication, that is,
for any integer \code{x}, \code{1 * x = x}.  (Imagine using \code{0}
for \code{y} instead. The result of \code{foldl} would be
uninteresting.)

Mathematicians have been discovering such abstractions for quite some
time, defining \emph{algebraic structures} to describe these
abstractions and proving reusable theorems about these
abstractions~\cite{van-der-Waerden:1990kn}. An algebraic structure
consists of: one or more sorts, signatures for operations over the
sorts, and axioms that describe the semantics of the operations.  For
example, the algebraic structure that matches the needs of
\code{foldl} is Monoid. A Monoid consists of a sort $S$ together with
a binary operation on $S$ that is associative and that has an identity
element in $S$. An \emph{instance} of an algebraic structure is a set
for each sort, and an operation for each signature, such that the
axioms are true. The set of integers with addition and \code{0} is an
instance of Monoid.

In the late 1970s and early 1980s,
\citet{Burstall:1977yf,Burstall:1980eu} and Kapur, Musser, and
Stepanov~\cite{Kapur:1981fx} noticed that it is helpful to think of
computer algorithms as operating on algebraic structures instead of
concrete data structures.  To quote Stepanov:
\begin{quote}
  {\it That is the fundamental point: algorithms are defined on algebraic
  structures~\cite{Russo:2011vn}.}
\end{quote}
In this setting, algebraic structures are analogous to interfaces (as
in Java) and instances are analogous classes that implement the
interfaces.

However, the analogy between algebraic structures and interfaces is
superficial because they differ both with regards to both purpose and
semantics.  Interfaces are primarily created to categorize classes
whereas algebraic structures are created to express the requirements
of theorems (in mathematics) and algorithms (in computer science).
With respect to semantics, a class may inherit from an interface if it
provides methods that satisfy the variance rules required by
\emph{subtyping} (contravariant parameter types and covariant return
types).  On the other hand, a class implements an algebraic data
structure if \emph{substituting} the class for the structure's sort in
the structure's operation signatures yields a set of signatures that
are implemented by the class.  This subtle difference is significant:
interfaces suffer from the binary method problem whereas algebraic
structures do not~\cite{Bruce:1995fk}.

The late 1970s and 1980s saw considerable developments both in the
practice of generic programming and in the design of language features
that support generic programming.
Kershenbaum, Musser, and Stepanov~\cite{Kershenbaum88} developed
generic sorting and searching algorithms in Scheme and then
\citet{Musser87} developed similar algorithms in Ada.  Recall that
Scheme is a dynamically typed language, so it is straightforward to
express polymorphic algorithms in Scheme.  Ada, on the other hand, was
one of the early statically typed languages to support parametric
polymorphism. However, neither Scheme or Ada provided support for
organizing parameters of generic algorithms. (This support was added
to Ada later, in 1995, with the addition of generic package
parameters.)

\subsection{Programming Language Support for Generic Programming}

There were several lines of research in the 1980s to support the
specification and use of algebraic structures in programming.
\citet{Burstall:1977yf,Burstall:1980eu} and \citet{Kapur:1982fk}
developed the specification languages CLEAR and Tecton, respectively.
An example of defining some algebraic structures and using them in a
generic \code{reduction} function (a generalization of \code{foldl}),
is shown in Figure~\ref{fig:tecton-reduction}, written in Tecton.
Burstall collaborated with MacQueen on the HOPE
language~\cite{Burstall:1980kx}, which in turn inspired MacQueen's
work on the \emph{signatures} and \emph{functors} of the Standard ML
module system~\cite{MacQueen:1984vn}. In the field of computer
algebra, \citet{Jenks:1981uq} created the Scratchpad language.
\citet{liskov79:_clu_ref}, while not explicitly targeting algebraic
structures, did include a feature called \emph{type set} in CLU that
could be used to express algebraic structures. CLU was meant to
support \emph{abstract data types}, but it turns out that abstract
data types and algebraic structures are closely related.

\begin{figure}[tbp]
  \centering
\begin{lstlisting}[morekeywords={provide,create,instantiate}]
  create semigroup(S : set, +: S$\times$S $\to$ S)
    with x + (y + z) = (x + y) + z;

  create monoid(S : semigroup, 0: () $\to$ S)
    with 0 + x = x + 0 = x;

  create sequence(S : set, E : set, isnull : S $\to$ bool, head : S $\to$ E, tail : S $\to$ S);

  provide sequence of monoid with
    reduction: 
     x $\to$ if isnull(x) then 0 else head(x) + reduction(tail(x))

  integers(I : set, + : I $\times$ I $\to$ I, * : I $\times$ I $\to$ I, 0 : () $\to$ I, 1 : () $\to$ I)

  instantiate monoid of integers (S=I, + = +, 0 = 0)
  instantiate monoid of integers (S=I, + = *, 0 = 1)
\end{lstlisting}
  \caption{A generic reduction function written in Tecton.}
  \label{fig:tecton-reduction}
\end{figure}

By the 1980s, object-oriented languages were gaining momentum, and
work was underway to integrate parametric polymorphism into
object-oriented languages. \citet{Cardelli:1985kx} developed
\emph{bounded polymorphism}, that is, using subtyping to express
constraints on type parameters. \citet{Canning:1989fk} generalized
bounded polymorphism to enable recursive bounds, which laid the
foundation for generics in Java~\cite{Bracha:1998fk} and
C\#~\cite{Kennedy:2001uq,Yu:2004kx}.


In 1988, \citet{stroustrup88:_param_types} added support for generic
programming to \cpp{} with the addition of templates. Stroustrup
considered using bounded polymorphism, but it was not a natural fit
for many of the use cases he had in
mind~\cite{stroustrup_concept_checking03}. Stepanov suggested using a
design similar to Ada's generic packages, with explicit instantiation,
but Stroustrup disagreed with such a design because he thought
explicit instantiation would be an unreasonable burden on clients of a
generic
libraries~\cite{Stroustrup:1993fk,Stevens:1995vn,Stroustrup:2007uq}.
Stroustrup also looked at the design choices in
ML~\cite{Stroustrup:1988kx}, but he did not model \cpp{} templates
after ML's functors for the same reason: functors required explicit
instantiation.  However, some confluence in the two designs can be
seen in \cpp{}'s template argument deduction for function templates.
The algorithm is based on the mathematical notion of \emph{matching},
which is the single-sided version of the \emph{unification} algorithm at
the heart of Hindley-Milner type inference.


Templates were designed to be safer than C preprocessor macros but
just as efficient~\cite{stroustrup88:_param_types}.  Macros had been
used for years as ``a poor man's generics''.  The design for type
checking templates followed a similar model to that of macros: type
checking occurs after instantiation.  (For templates, some
checking happens before instantiation, but not much.)  This design
choice had some significant advantages and disadvantages.  The primary
advantages were extreme flexibility and expressiveness: it opened the
door to template
metaprogramming~\cite{Veldhuizen:1995kx,Alexandrescu:2001kl,Abrahams:2004oq}.
The primary disadvantage was the lack of modular type checking.  We
discuss the semantics and type system for templates in more detail in
Section~\ref{generic-programming}.  The flexibility of templates,
combined with function overloading, provided a means to write
type-parameterized functions and to implicitly parameterize over
operations, thereby making it relatively convenient to write generic
algorithms in \cpp{}.  In concert with Stroustrup's work on templates,
Stepanov began developing a \cpp{} component
library~\cite{Stepanov:1991ys} along the lines of his prior work in
Scheme and Ada. This \cpp{} library was the direct precursor to the
Standard Template Library.

The 1980s ended with an important result: \citet{Kaes:1988fk} and
\citet{wadler89:_adhoc} discovered parametric
overloading, then \citet{wadler89:_adhoc} expanded the idea to create
the \emph{type class} feature for Haskell. Type classes were not
explicitly designed to support algebraic structures. Instead they were
a generalization of ML's equality types. Nevertheless, type classes
provide excellent support for generic programming, combining a modular
type system with the convenience of implicit instantiation of generics
(in contrast to the explicit functor application of Standard ML). If
only the divide between the imperative and functional programming
communities had not been so large!




\subsection{Road Map}

In the next section we briefly survey the style of generic programming
that is used to develop modern \cpp{} template libraries, as
\emph{concepts} are meant to support this style of programming.  After
that, we discuss the preliminary language design work at Texas and
Indiana (Section~\ref{sec:prelim}).  We then take a look at the two
proposals to the \cpp{} standards committee, first the Indiana
proposal (Section~\ref{sec:indiana-proposal}) and then the Texas
rebuttal (Section~\ref{sec:texas-rebuttal}). The two teams formed a
compromise at the Adobe meeting (Section~\ref{sec:adobe}) which lead
to the final design for concepts that was voted into the \cpp{}
standard working draft (Section~\ref{sec:acceptance}). Unfortunately,
the compromise unraveled and concepts were voted back out of \cpp{}
(Section~\ref{sec:removal}), leaving much in doubt regarding the
future of concepts in \cpp{} (Section~\ref{sec:conclusion}).

\section{Generic Programming and \CPP{} Templates}
\label{generic-programming}

\lstset{style=cpp}

In this section we take a closer look at the modern practice of
generic programming in \cpp{}. The examples are from the Standard
Template Library~\cite{stepa.lee-1994:the.s:TR}, though they could
have easily come from many other generic libraries. The presentation
here is in the style of the SGI version of the
STL~\cite{austern99:_gener_progr_stl}.

In the parlance of modern \cpp{}, \emph{concept} means algebraic
structure and \emph{model} means an instance of an algebraic
structure~\cite{Stevens:1995vn,austern99:_gener_progr_stl}.  The shift
away from using the term ``algebraic structure'' makes sense because
many of the abstractions have little to do with algebra. Also, the
term ``structure'' is too close to \code{struct}, which has a very
different meaning in \cpp. Unfortunately, the term ``concept'' is
rather generic.

\begin{figure}[htbp]
  \centering
  \includegraphics[width=4in]{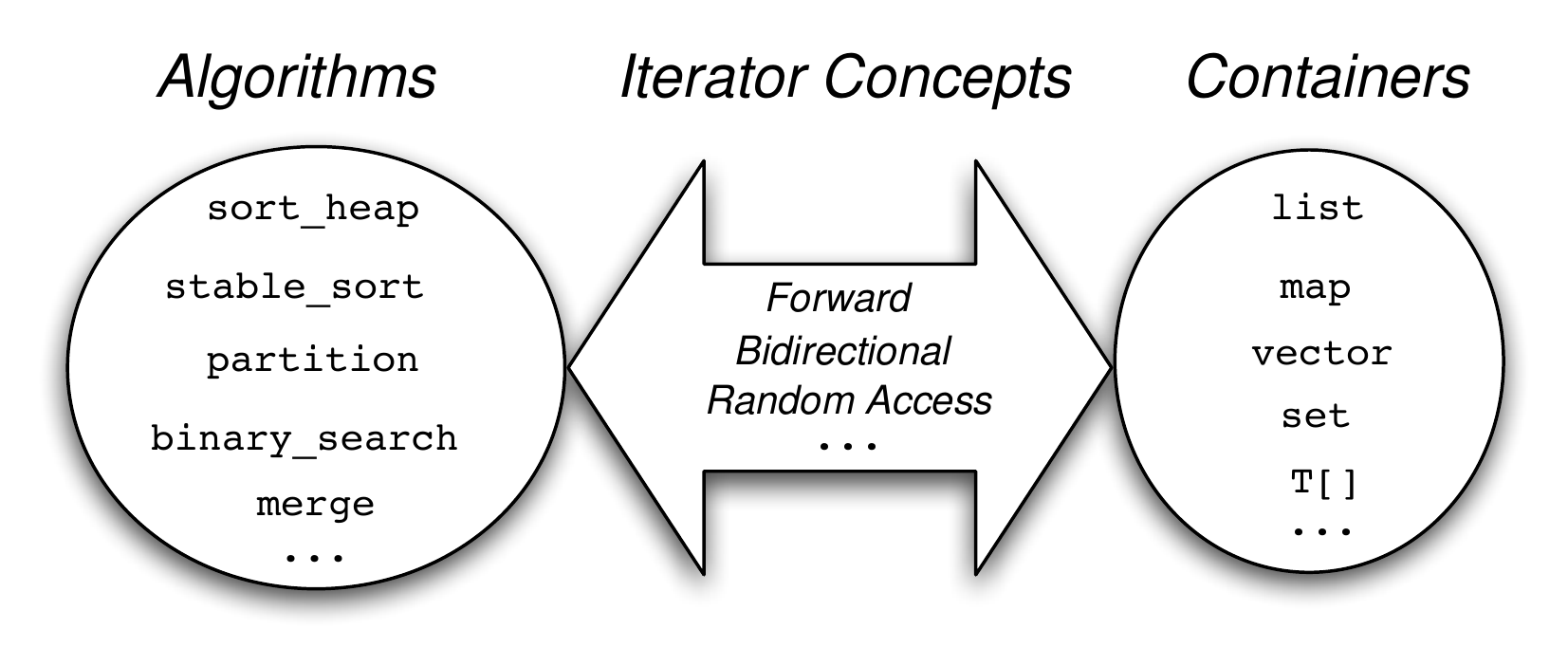}  
  \caption{Organization of the Standard Template Library.}
  \label{fig:algo-iter-data}
\end{figure}

Figure~\ref{fig:algo-iter-data} depicts the organization of the
Standard Template Library. The STL contains 112 generic algorithms and
12 container classes. More importantly, the STL provides a handful of
iterator concepts that decouple the algorithms from the containers.
Because of the iterator abstractions, there only needs to be 112
algorithm implementations and 12 iterator implementations to enable
the algorithms to work with all of the containers. (At least, for all
the combinations that makes sense.) Using traditional procedural or
object-oriented techniques, there would be $112 \times 12$ algorithm
implementations, as each algorithm is implemented by different code
for each container class.

\begin{figure}[htbp]
  \centering
\includegraphics[width=5in]{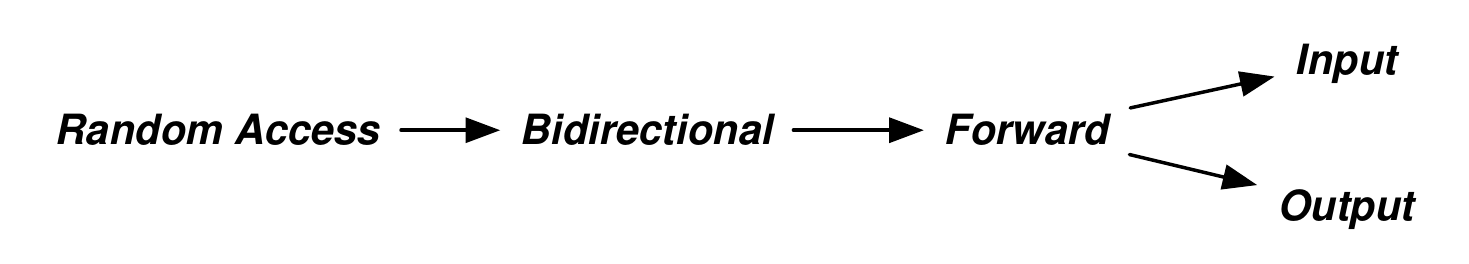}
  \caption{The Iterator Concepts of the Standard Template Library.}
  \label{fig:iterator-concepts}
\end{figure}

Figure~\ref{fig:iterator-concepts} depicts the hierarchy of iterator
concepts within the STL. The arrows indicate the \emph{refinement}
relation between concepts (analogous to inheritance between
interfaces). A concept refines another concept if it includes all of
the capabilities of the other concept. Or put another way, concept
$C_1$ refines $C_2$ if all the models of $C_1$ are also models of
$C_2$. The Random Access Iterator concept requires the most
capabilities, in particular it requires the capability of jumping
forward or backwards an arbitrary number of positions in constant
time. The Input and Output Iterator concepts require the least
capabilities, just requiring the ability to make a single pass through
a sequence, respectively reading or writing to the elements.

The iterator hierarchy provides a fine-grained categorization of
iterator capabilities to enable each algorithm to precisely specify
its requirements. The precision is necessary to maximize reusability:
each algorithm should use the minimal requirements necessary to
efficiently carry out its task, thereby maximizing the number of
situations in which it can be used.
An important point to stress is that, in generic programming, the
raison d'\^etre for concepts is to provide concise means for
algorithms to specify their requirements. This viewpoint contrasts
with object-oriented methodologies in which interfaces are created to
categorize classes independently of their use in algorithms.

Figure~\ref{fig:input-iterator} shows the description of the Input
Iterator concept from the SGI STL~\cite{austern99:_gener_progr_stl}.
In the description, the type variable \code{X} is used as a place
holder for the modeling type.
The valid expressions specify the operations that must be provided by
the modeling type. The complexity guarantees enable generic algorithms
to in turn state guarantees regarding their execution time.
Associated types are types that have an auxiliary role in the concept
and whose identity depends on (is a function of) the modeling type.
(Associated types are related to virtual
types~\cite{Bruce:1998ua,torgersen98:_virtual_types_safe,ernst99b,Odersky:2003on},
but avoid many of the difficulties by being accessed through types
instead of objects. Associated types are most closely related to
abstract types of a signature in ML.) In \cpp{}, associated types are
usually accessed through a template trick known as
traits~\cite{myers95:_trait}.

\begin{figure}[tbp]
\textbf{Description}
\vspace{2pt}

An input iterator provides the capability of traversing through a
sequence once and reading the values of the elements.
\vspace{5pt}

\textbf{Associated Types}
\vspace{-3pt}
\begin{itemize}\addtolength{\itemsep}{-0.5\baselineskip}
\item \code{iterator_traits<X>::value_type}, this is the element
  type of the sequence.
\item \code{iterator_traits<X>::difference_type}, this is for
  measuring distances between positions in the sequence.
\end{itemize}

\textbf{Valid Expressions} 
\vspace{2pt}

  (\code{X} is the iterator type, \code{T} is the value type)\\
{\footnotesize
\begin{tabular}[c]{{|l|l|p{2in}|}} \hline
\textbf{expression} & \textbf{return type} & \textbf{semantics} \\ \hline\hline
\code{*i} & Convertible to \code{T} & Returns the value at position \texttt{i} \\ \hline
\code{++i} & \code{X&} &  Moves the iterator to the next position. \\ \hline
\code{i == j} & \code{bool} & Returns true if \code{i} and \code{j}
  are at the same position. 
  \code{i == j} implies \code{*i == *j}
\\ \hline
\code{i != j} & \code{bool} & Equivalent to \code{!(i == j)}. \\ \hline
\end{tabular}
}

\vspace{10pt}
\textbf{Complexity guarantees} 
\vspace{2pt}

  All operations are amortized constant time.
  \caption{The Input Iterator Concept}
  \label{fig:input-iterator}
\end{figure}

Figure~\ref{fig:merge} lists the \code{merge} function template from
the STL, which serves here as a typical example of a generic
algorithm. The documentation for \code{merge} states that the type
parameters \code{InIter1} and \code{InIter2} must model Input
Iterator. The need for this requirement can be seen in that the
algorithm increments, dereferences, and uses the not-equal operator
on the the iterators \code{first1}, \code{last1}, \code{first2}, and
\code{last2}. The \code{OutIter} type is required to model Output
Iterator, which enables the dereference and assignment to
\code{result} as well as the incrementing of \code{result}.  Also, to
enable the assignment, the value type of \code{OutIter} must be the
same as the value types of \code{InIter1} and \code{InIter2}.  And
finally, the if-statement compares the elements of the two input
iterators, so the associated value type of \code{InIter1} must model
the Less Than Comparable concept. The \code{merge} function calls the
\code{copy} function template, which is also part of the STL.  The
\code{copy} template in turn places requirements on its type
parameters, and those requirements are a subset of the type
requirements for \code{merge}.

\begin{figure}[tbp]
  \centering
\begin{lstlisting}
template <typename InIter1, typename InIter2, typename OutIter>
OutIter
merge(InIter1 first1, InIter1 last1, InIter2 first2, InIter2 last2, OutIter result) {
  while (first1 != last1 && first2 != last2) {
    if (*first2 < *first1) {
      *result = *first2;
      ++first2;
    } else {
      *result = *first1;
      ++first1;
    }
    ++result;
  }
  return copy(first2, last2, copy(first1, last1, result));
}  
\end{lstlisting}
  \caption{The Merge Function Template of the STL.}
  \label{fig:merge}
\end{figure}


\subsection{The Semantics of \cpp{} Templates}

The main idea behind the semantics of templates is that different
versions of a template are stamped out for different uses.  The
\code{min} function template and its use in \code{main} in
Figure~\ref{fig:min}, serves to demonstrate the key ideas.  At the
point where \code{min} is defined, a \cpp{} type checker looks at any
expressions in the body of \code{min} whose type does not depend on
the type parameters.  In this case, there are none, so no type
checking occurs at the point of definition of \code{min}.

\begin{figure}[tbp]
  \centering
\begin{lstlisting}
namespace std {
  template <class T>
  T min(T a, T b) {
    if (b < a) return b; else return a;
  }
}
int main() {
  return std::min(3, 4);
}
\end{lstlisting}
  \caption{A simple function template: \code{the} min operation.}
  \label{fig:min}
\end{figure}

Moving on to the \code{main} function, the call to \code{std::min} is
an example of template argument deduction.  A \cpp{} compiler
deduces that \code{int} should be chosen for parameter \code{T} by
pattern matching the argument types against the parameter types. So
the compiler transforms the call to include an explicit instantiation:
\begin{lstlisting}
int main() {
  return std::min<int>(3, 4);
}
\end{lstlisting}
A \cpp{} compiler then generates a version of \code{std::min}
that is specialized for \code{int}. The following is a source code
representation of the generated code (which usually only exists
in the internal representation of the compiler).
\begin{lstlisting}
namespace std {
  template<>
  int min<int>(int a, int b) {
    if (b < a) return b; else return a;
  }
}
\end{lstlisting}
A \cpp{} compiler type checks the specialized code, which
in this case is well typed. If we change the example to apply
\code{min} to a type that does not provide a less-than comparison
operator, as follows,
\begin{lstlisting}
struct A {};
int main() {
  A a;
  return std::min(a, a);
}
\end{lstlisting}
we get an error message that points inside the \code{min} function
template.
\begin{lstlisting}[deletekeywords=for]
error1.cpp: In function 'T std::min(T, T) [with T = A]':
error1.cpp:8:   instantiated from here
error1.cpp:3: error: no match for 'operator<' in 'b < a'
\end{lstlisting}
For such a small template, this error is not difficult to debug, but
for most templates in the STL, such errors are infamously difficult
to comprehend. For example, the following program that mistakenly
applies \code{stable_sort} to a linked list results in the error
message in Figure~\ref{fig:infamous-error}. Can you see the problem?
\begin{lstlisting}
int main() {
  list<int> l;
  stable_sort(l.begin(), l.end());
}
\end{lstlisting}
The \code{stable_sort} template requires the iterators to provide
random access, but the iterators of a double linked list are
merely bidirectional.

These examples demonstrate that \cpp{} does not have a modular type
system. When checking the use of a template, the \cpp{} type checker
looks at the body of the template. To be modular, \cpp{} would need
the notion of the type of a template, like the universal type of
System F, and it would have to check uses of templates against such
types, but it does not.

As more programmers began to use the STL in the late 1990s and early
2000s, such error messages became a constant aggravation and deterred
many programmers from using the STL.  \citet{Meyers:2001fk}, an author
of many popular \cpp{} books, writes
\begin{quote}
  \emph{Perhaps most daunting, even the smallest STL usage error often
    led to a blizzard of compiler diagnostics, each thousands of
    characters long, most referring to classes, functions, or
    templates not mentioned in the offending source code, almost all
    incomprehensible.}
\end{quote}

Building on an idea of Stepanov's, I developed a \cpp{} library to
check requirements on type parameters and thereby improve the error
messages~\cite{Siek:2000if}. However, using the library required
writing tedious and error prone code, and the quality of the error
messages was dependent on the compiler.

\begin{figure}[tbp]
  \centering
\begin{lstlisting}[basicstyle=\tiny\sffamily,lineskip=-2pt,deletekeywords=for]
stl_algo.h: In function 'void std::__inplace_stable_sort(_RandomAccessIterator, _RandomAccessIterator)
  [with _RandomAccessIterator = std::_List_iterator<int>]':
stl_algo.h:3633:   instantiated from 'void std::stable_sort(_RandomAccessIterator, _RandomAccessIterator) 
  [with _RandomAccessIterator = std::_List_iterator<int>]'
error1.cpp:8:   instantiated from here
stl_algo.h:2921: error: no match for 'operator-' in '__last - __first'
stl_algo.h:3633:   instantiated from 'void std::stable_sort(_RandomAccessIterator, _RandomAccessIterator) 
  [with _RandomAccessIterator = std::_List_iterator<int>]'
error1.cpp:8:   instantiated from here
stl_algo.h:2926: error: no match for 'operator-' in '__last - __first'
stl_algo.h:2929: error: no match for 'operator-' in '__middle - __first'
stl_algo.h:2929: error: no match for 'operator-' in '__last - __middle'
stl_algo.h: In function 'void std::__stable_sort_adaptive(_RandomAccessIterator, _RandomAccessIterator, _Pointer, _Distance) 
  [with _RandomAccessIterator = std::_List_iterator<int>, _Pointer = int*, _Distance = int]':
stl_algo.h:3635:   instantiated from 'void std::stable_sort(_RandomAccessIterator, _RandomAccessIterator) 
  [with _RandomAccessIterator = std::_List_iterator<int>]'
error1.cpp:8:   instantiated from here
stl_algo.h:3550: error: no match for 'operator-' in '__last - __first'
stl_algo.h:3551: error: no match for 'operator+' in '__first + __len'
stl_algo.h:3635:   instantiated from 'void std::stable_sort(_RandomAccessIterator, _RandomAccessIterator) 
  [with _RandomAccessIterator = std::_List_iterator<int>]'
error1.cpp:8:   instantiated from here
stl_algo.h:3564: error: no match for 'operator-' in '__middle - __first'
stl_algo.h:3564: error: no match for 'operator-' in '__last - __middle'
stl_algo.h: In function 'void std::__insertion_sort(_RandomAccessIterator, _RandomAccessIterator) 
  [with _RandomAccessIterator = std::_List_iterator<int>]':
stl_algo.h:2923:   instantiated from 'void std::__inplace_stable_sort(_RandomAccessIterator, _RandomAccessIterator)
  [with _RandomAccessIterator = std::_List_iterator<int>]'
stl_algo.h:3633:   instantiated from 'void std::stable_sort(_RandomAccessIterator, _RandomAccessIterator) 
  [with _RandomAccessIterator = std::_List_iterator<int>]'
error1.cpp:8:   instantiated from here
...
\end{lstlisting}
  \caption{The prefix of a typical error message that arises from misusing the STL.}
  \label{fig:infamous-error}
\end{figure}

\subsection{Templates and Separate Compilation}
\label{sec:template-spec-separate-comp}

The specialization-based semantics of \cpp{} templates rules out the
possibility of separate compilation.  Other languages, such as ML and
Java provide generics but retain separate compilation, so why can't
\cpp{}?  It turns out that not only is specialization the main
implementation approach for compiling templates, but the semantics of
the language forces it to be the only implementation approach. 

One might think that a \cpp{} compiler could support both separate
compilation and specialization by postponing template specialization
to run time.  However, the reader may recall that the definition of
separate compilation (Section~\ref{sec:higher-order-poly}) requires
that the amount of time used for compilation (whether it be at compile
time or run time) be proportional to the size of the application, not
including the size of the (template) libraries that it uses. Thus,
even an implementation using run-time specialization would not provide
separate compilation.

\begin{figure}[tbp]
  \centering
\begin{lstlisting}
template<typename T>
struct C {
  typedef double num;
};

template<>
struct C<char> {
  typedef int num;
};

template<typename T>
void f(T x) {
  typedef typename C<T>::num U;
  U A[2];
  A[sizeof(U)/4 $-$ 1] = x;
}

int main() {
  f('z');
  f(42);
  f(pair(1,2));
}
\end{lstlisting}
  \caption{User-defined specialization inhibits separate compilation.}
  \label{fig:sep-comp}
\end{figure}

There are several language features in \cpp{} that force
specialization; the one we describe here is user-defined template
specialization.  The example in Figure~\ref{fig:sep-comp} demonstrates
that many phases of a \cpp{} compiler, from front-end type checking to
back-end code generation, require specialization. The example shows a
function template \code{f} that refers to class template \code{C},
using its nested type \code{num} to declare an array \code{A}.
The call \code{f('z')} causes \code{C} to be instantiated with the
type argument \code{char}, which results in the user-defined
specialization of \code{C} and the typedef of \code{num} to
\code{int}. So \code{sizeof(U)} is 4 bytes (assuming a 32 bit
architecture) and therefore \code{sizeof(U)/4 $-$ 1} is zero, causing
\code{x} to be assigned to the first element of the array.  The call
\code{f(42)} causes \code{C} to be instantiated with \code{int}, which
results in the instantiation of the primary template and the typedef
of \code{num} to \code{double}. So \code{sizeof(U)} is 8 bytes and
therefore \code{sizeof(U)/4 $-$ 1} is one, causing \code{x} to be
assigned to the second element of the array.  The call
\code{f(pair(1,2))} results in a type error because \code{pair} is not
convertible to \code{double}.

This example shows that \cpp{} templates are not a form of parametric
polymorphism (as technically defined) because the behavior of a
template may depend on its type arguments.
But more importantly, it shows that compilation time is a function of
the number of template instantiations.
For example, the type checking of \code{f}'s body must be repeated for
each unique instantiation because there can be a different outcome for
each one: \code{f(42)} was well typed but not \code{f(pair(1,2))}.
Furthermore, the generated assembly code for the instantiations of
\code{f} would need to differ in many ways.  For example, most
architectures provide different instructions for storing \code{int}
versus \code{double}, so the assignment of \code{x} into \code{A}
would require different instructions.
Thus, in general, a \cpp{} compiler needs to generate different code
sequences for different instantiations.

%
So the compilation time for a \cpp{} application must be a function of
the size of all the templates libraries it uses, transitively.  For
modern \cpp{} applications, the compilation time can stretch from
minutes into hours.

\subsection{Templates and Name Lookup}

While non-modular error messages are the most obvious problem with
the semantics of \cpp{} templates, there are also subtle problems
regarding name lookup. Consider the \code{min} template again, but
this time we apply it to a user-defined type that provides a
less-than operator.
\begin{lstlisting}
namespace L {
  class interval { };
  bool operator<(interval x, interval y) { ... }
}
int main() {
  L::interval i, j, k;
  k = std::min(i,j);
}
\end{lstlisting}
The above example is well typed, but how? How does the use of the
less-than operator inside \code{std::min} resolve to the less-than
operator in namespace \code{L}?  The answer is \emph{argument
  dependent name lookup} (ADL).  When a \cpp{} compiler performs name
lookup for the function name in a function call, it not only considers
the names that are in lexical scope but it also analyzes the argument
types and then includes all the names in the namespaces in which those
types were defined. In this case, the less-than operator is applied to
arguments of type \code{interval}, and \code{interval} was defined in
namespace \code{L}. So the less-than operator in namespace \code{L} is
among the entities returned by name lookup in this case.

\begin{figure}[tbp]
  \centering
\begin{lstlisting}
namespace lib {
  template <class T> void load(T, string) { printf("Proceeding as normal!\n"); }
  template<class T> void initialize(T x) { load(x, "file"); }
}
namespace N {
  struct b { int n; };
  template <class T> void load(T, const char*) { printf("Hijacked!\n"); }
  template<class T> void shoot(T x) { load(x, "ammunition"); }
}
int main() {
  N::b a;
  lib::initialize(a);
}
// Output: Hijacked!
\end{lstlisting}
  
  \caption{Argument dependent lookup can lead to unintended run-time behavior.}
  \label{fig:adl-bad}
\end{figure}

While argument dependent name lookup may seem like a good thing, it is
not modular and causes bugs when building large software systems. The
example in Figure~\ref{fig:adl-bad} is distilled from a real bug that
showed up in an application using the Boost \cpp{} libraries.  The
\code{initialize} function template in namespace \code{lib} intends to call
another function template in \code{lib} named \code{load}. Unbeknownst
to the author of \code{lib}, there is another function named
\code{load} in namespace \code{N} that is meant to be a helper
function for \code{shoot}.  Because of argument dependent lookup, the
call to \code{load} inside of \code{initialize} does not resolve to
\code{lib::load}, but instead resolves to \code{N::load}. Thus, the
output of the above program is ``Hijacked!''. What is particularly
troubling about this example is that the run-time behavior of
\code{lib::initialize} depends on the context in which it is used,
analogous to the infamous dynamic scoping of
Lisp~\cite{Guy-L.-Steele:1982bj}.

\subsection{Tag Dispatching}

On the positive side, the semantics of \cpp{} templates enables some
powerful programming styles. For example, there is a natural tension
between performance and generality in algorithm design. Often times
there exist multiple algorithms to solve the same problem, where one
algorithm is more efficient than another, but requires more
capabilities from the types it operates on. The \code{advance}
function of the STL is a particularly simple example of this. The
\code{advance} function moves an iterator forward $n$ positions within
the sequence.  If the iterator models Random Access Iterator, then the
iterator can directly jump forward $n$ positions in constant time. On
the other hand, if an iterator only models Input Iterator, then it
takes linear time to move forward $n$ positions. 

An obvious solution to this problem is to have two differently named
functions for advancing iterators. However, what if the need to
advance iterators appears in the context of a generic algorithm?  For
example, the \code{lower_bound} function template of the STL calls
\code{advance}, but is itself parameterized on the iterator types.  In
that context we do not know which iterators are being used.
The solution to this problem in \cpp{} is the tag dispatching idiom,
shown in
Figure~\ref{fig:tag-dispatch}\footnote{\citet{Kiselyov:2011fk}
  describe a solution to this problem in Haskell.}.  We create a
single entry point named \code{advance} and then use traits to inquire
about the capability of the iterator. The resulting tag is used to
influence the function overload resolution for the call to the helper
functions named \code{advance_aux}.  The tag dispatching trick works
because the overload resolution of the call to \code{advance_aux}
inside of \code{advance} does not occur until after \code{advance} has
been instantiated on particular concrete iterator types.

\begin{figure}[tbp]
  \centering
\begin{lstlisting}
template <class InIter, class Dist>
void advance_aux(InIter& i, Dist n, input_iterator_tag)
  { while (n--) ++i; }

template <class RandIter, class Dist>
void advance_aux(RandIter& i, Dist n, random_access_iterator_tag)
  { i += n; }

template <class InIter, class Dist>
void advance(InIter& i, Dist n) {
  typename iterator_traits<InIter>::iterator_category cat;
  advance_aux(i, n, cat);
}
\end{lstlisting}
  \caption{An example of the tag dispatching idiom.}
  \label{fig:tag-dispatch}
\end{figure}
\subsection{Evaluation of \cpp{} Templates}

To briefly evaluate the design of \cpp{} templates, on the positive
side \cpp{} effectively enables generic programming by providing type
parameterization and the convenient use of function templates through
template argument deduction and argument dependent lookup.
Furthermore, the run-time performance of \cpp{} templates can match
that of hand-coded procedures because the specialization approach leaves
behind little or no run-time overhead. Last but not least, the lack of
point-of-definition type checking for templates provided the
flexibility needed for template metaprogramming.

On the negative side, the type checking of templates is not modular
which leads to confusing error messages for users of generic libraries
and opens the door to discrepancies between a library's documentation
and implementation. Further, because argument dependent lookup is
non-modular, the run-time behavior of template libraries can silently
change in unexpected ways in different usage scenarios.  Finally, the
specialization semantics of \cpp{} templates prevents separate
compilation, so developers of large \cpp{} applications endure long
compilation times.


\section{Preliminary Research on Concepts}
\label{sec:prelim}

In 2003, two teams set out to design an extension to \cpp{} to improve
\cpp{} templates by providing modular type checking. To type check the
body of a template independently of any instantiation, the type
checker needs to know what assumptions are being made about the type
parameters. As discussed in the previous section, these assumptions
were commonly stated in the documentation for the template; what was
needed was a way to make these assumptions explicit in the code.  The
two teams set out to add support for type constraints, using concepts
to express the constraints. At Texas A\&M University, Bjarne
Stroustrup (having just moved from AT\&T Labs--Research) and then
post-doc Gabriel Dos Reis began the Pivot project and produced several
technical reports for the \cpp{} standards committee.  We discuss
these reports in Section~\ref{sec:kona}.  

At Indiana University, Professor Andrew Lumsdaine, post-doc Jaakko
J\"arvi, and the graduate students Ronald Garcia, Jeremiah Willcock, and
I also began working on concepts and constrained generics.  We
did not have prior experience in language design, but we had
considerable experience in the design and implementation of generic
libraries~\cite{Jarvi:2001fk,Siek:2002ly,Siek:1999lf,Lee:1999ry,Siek:2000ca,Jarvi:2003uq,Garcia:2005ve}.
Our first major undertaking was to study support for generic
programming in state of the art programming languages, discussed in
the following section.


\subsection{Indiana: A Comparative Study}
\label{sec:comparative}

In our study~\cite{Garcia:2003gn} we implemented several generic
algorithms from the Boost Graph Library~\cite{Siek:2002ly} in six
languages: \cpp{}, Standard ML, Haskell, Eiffel, Java, and C\#. (In a
follow-on study, we added OCaml and Cecil~\cite{Garcia:2007fk}.)  We
sought to use best practices for each language. In the course of the
study, we took note of when the presence or absence of a language
feature made the implementation more or less difficult.
Table~\ref{tab:study} summarizes the results of the study.

\input{comparison_chart}

The following explains the language features that form the rows of the
table.  ``Multi-type concepts'' indicates whether multiple types can
be simultaneously constrained. ``Multiple constraints'' indicates
whether more than one constraint can be placed on a single type
parameter.  ``Associated type access'' rates the ease in which types
can be mapped to other types within the context of a generic function.
``Retroactive modeling'' indicates the ability to add new modeling
relationships after a type has been defined. ``Type aliases''
indicates whether a mechanism for creating shorter names for types is
provided.  ``Separate compilation'' indicates whether generic
functions are type-checked and compiled independently from their use.
``Template arg. deduction'' indicates that type arguments can be
deduced without requiring explicit syntax for instantiation.
``Concise syntax'' indicates whether the syntax required to compose
layers of generic components is independent of the scale of
composition.

Haskell faired particularly well in this study, with Standard ML not
too far behind, while Eiffel, Java, and C\# did not do as well.  The
underlying pattern was that the three object-oriented languages relied
on F-bounded polymorphism~\cite{Canning:1989fk}, whereas Standard ML
and Haskell did not. Standard ML supports generic programming through
Functors and signatures and Haskell supports generic programming via
type classes. The take-away point for us was that a design for
concepts in \cpp{} should be based on the best features of Haskell and
Standard ML, and not F-bounded polymorphism. Our case study influenced
other programming language researchers.  For example,
\citet{Chakravarty:2005tf,Chakravarty:2005fl} added associated types
to Haskell, filling in the only half-circle for Haskell in
Table~\ref{tab:study}.

\subsection{Texas: Proposal at Kona}
\label{sec:kona}

At the October, 2003 \cpp{} standards committee meeting in Kona,
Stroustrup and Dos Reis presented three technical reports,
N1510~\cite{stroustrup_concept_checking03},
N1522~\cite{Stroustrup:2003vl}, and N1536~\cite{Stroustrup:2003yf}
that sketched a design for concepts.

There were five important aspects to the Texas design. First, they
proposed \emph{usage patterns} as the mechanism for specifying which
operations are required of the modeling type, that is, the type
implementing the concept. Usage patterns mimic the notation that had
become common for documenting generic libraries.  The following
example shows a concept that requires the modeling type to provide a
copy constructor, an addition operator, and an assignment operator.
\begin{lstlisting}
concept Add {
  constraints(Add x) { Add y = x; x = x+y; } 
};   
\end{lstlisting}

Second, the design proposed \emph{composing concepts} using the logical
operators \code{&&} (and), \code{||} (or), and \code{!} (not).  In the
following, the concepts \code{C1} and \code{C2} are combined in
different ways to constrain the \code{T} type parameter of three
classes.
\begin{lstlisting}
template <(C1 && C2) T> class X { ... }; 
template <(C1 || C2) T> class Y { ... }; 
template <(C1 && !C2) T> class Z { ... };  
\end{lstlisting}

Third, the proposal did not include support for associated types, but
instead proposed \emph{parameterized concepts}. In the following
\code{Forward_iterator} concept, the value type of the iterator is a
type parameter.
\begin{lstlisting}
template <Value_type V> concept Forward_iterator { 
  constraints(Forward_iterator p) { 
    Forward_iterator q = p; V v = *p;  p++; ++p;
  } 
}   
\end{lstlisting}

Fourth, the proposal included \emph{concept-based overloading}.  That
is, extending the rules for function overload resolution to take into
account the constraints on type parameters.  This feature was meant to
replace the tag dispatching idiom.  For example, below are the two
\code{advance} algorithms, the first requiring only an Input Iterator
and the second requiring a Random Access Iterator.
\begin{lstlisting}
template <InputIterator InIter>
void advance(InIter& i, InIter::difference_type n)
  { while (n--) ++i; }

template <RandomAccessIterator RandIter>
void advance(RandIter& i, RandIter::difference_type n)
  { i += n; }
\end{lstlisting}

Fifth and finally, the proposal used \emph{implicit modeling}.  That
is, a \cpp{} compiler would automatically deduce whether a class
models a concept. In the following example, we instantiate the
\code{f} function template with class \code{A}. Template \code{f}
requires \code{A} to model the \code{Add} concept, so the \cpp{}
compiler would check that \code{A} implements all of the operations
required by the \code{Add} concept.

\begin{lstlisting}
class A { };
A operator+(A x, A y) { ... }

template<Add T> void f(T x) { ... }

int main() {
  A a;
  f(a); // compiler deduces that A models Add
}
\end{lstlisting}

At the Kona meeting and in the following months, several problems with
the proposal were identified. With respect to usage patterns, there
was an open question regarding implementation.  To type check the body
of a template, the usage patterns would need to be converted to type
signatures for use in the typing environment (that is, the symbol
table).  Supposing that hurdle could be overcome, there was also an
issue regarding usability. For example, it seems that the following
template \code{h}, which seems fine, would in fact be ill typed.
\begin{lstlisting}
template<Add T> void g(T x, T y) { ... }

template<Add T> void h(T x, T y) {
  g(x, x + y); // error: no matching function for call to 'g(T, T1)'
}
\end{lstlisting}
To explain, the usage patterns for the \code{Add} concept, defined
above, would imply that, given a type \code{T} that models \code{Add},
addition would have the following signature:
\begin{lstlisting}
  T1 operator+(T,T);
  where T1 is some type convertible to T
\end{lstlisting}
Thus, in the call to \code{g}, the types of the two arguments would be
\code{T} and \code{T1}, respectively. But \code{g} requires that the
two arguments be of the same type.

The third issue regarding usage patterns was that any \cpp{}
expression was allowed as a usage pattern. Together with implicit
modeling and concept-based overloading, the \cpp{} compiler would need
the ability backtrack out of arbitrarily deep chains of template
instantiations, which would require significant re-engineering of the
industry's \cpp{} compilers.

With respect to composing concepts, the \code{&&} operator was
straightforward and expected, but the \code{||} and \code{!} operators
were both unexpected and problematic. They were unexpected because
their was no use of them in the documentation of \cpp{}
template libraries. The \code{||} operator was problematic because it seemed
that adding this operator would cause type checking to
become exponential. The \code{!} operator was problematic because its
semantics was unclear: how does the type checker make use of such a
constraint inside the body of a template? 
And on the instantiation
side, the \code{!} operator might require the close-world assumption
and therefore be incompatible with separate type checking.

With respect to parameterized concepts, the Texas design was similar
to Haskell's type classes, which supported multi-parameter type
classes but not associated types. However, our comparative study
showed that associated types provide a nice mechanism for reducing the
number of type parameters needed in generic
algorithms~\cite{Garcia:2003gn,Garcia:2007fk}.

Finally, the combination of implicit modeling and concept-based
overloading produces some dangerous
situations~\cite{gregor05:explicit_models}. Consider the \code{vector}
class from the STL shown in Figure~\ref{fig:implicit-overload}. It has
two constructors that accept an iterator range. The first only
requires the iterators to model Input Iterator, so the algorithm can
only make a single pass through the iterator range and must continually
resize the vector's memory to fit the incoming elements. The second
constructor requires the iterators to model Forward Iterator, so the
algorithm makes a first pass through the iterator range to determine
its length, then resizes the vector's memory to the appropriate size, and
finally makes a second pass through the iterator range, filling in the
elements of the vector.

\begin{figure}[tbp]
  \centering
\begin{lstlisting}
template<typename T>
class vector {
  template<InputIterator InIter> vector(InIter first, InIter last);
  template<ForwardIterator FwdIter> vector(FwdIter first, FwdIter last);
  ...
};

int main() {
  istream_iterator<int> i(cin), j;
  vector<int> v(i, j); // Silently dispatches to the wrong constructor!
}
\end{lstlisting}
  \caption{Implicit modeling and concept-based overloading don't mix.}
  \label{fig:implicit-overload}
\end{figure}

Consider the initialization of a \code{vector} from an
\code{istream_iterator<int>} (an iterator over integers from the
standard input). To determine which constructor to call, the \cpp{}
compiler would need to deduce whether \code{istream_iterator}
implements Forward Iterator and Input Iterator. Based on the syntactic
constraints captured in the concept definitions,
\code{istream_iterator} appears to model both concepts, so overload
resolution would choose the constructor for the more refined concept,
Forward Iterator (Figure~\ref{fig:iterator-concepts}). 
However, \code{istream_iterator} does not, in fact, model Forward
Iterator because it lacks the ability to make multiple passes through
its range.
Thus, the run-time behavior would be rather surprising: instead of
filling the vector with the first $n$ integers from standard input, it
would use the $n+1$ through $2n$ integers to fill in the vector
(supposing there is that much input available).

The ability to make multiple passes through a range is an example of a
\emph{semantic constraint}. Another example of a semantic constraint,
from the Equality Comparable concept, is that equality
should be reflexive, symmetric, and transitive, that is, an
equivalence relation. The Equality Comparable concept, from the
compiler's view, only requires that a type implement \code{operator==}
, but does not check that the \code{operator==} is actually an
equivalence relation.  To fully support semantic constraints requires
a highly expressive logic and support for theorem proving. For
example, the Isabelle proof assistant supports semantic constraints
with its axiomatic type classes~\cite{Wenzel:2005qy}. However, there
are major research and educational challenges that need to be resolved
before we can transplant those ideas into an mainstream, imperative
language such as \cpp{}. Thus, the concepts design did not include
semantic constraints, only syntactic ones. The recommended practice is
to document semantic constraints in English, as is done for the
Standard Template Library.

\subsection{Indiana: Two Designs}
\label{sec:kona}

Meanwhile Jaakko J\"arvi, Jeremiah Willcock, and I began designing
a prototype language with concepts and constrained templates.  Before
we proceeded very far, we realized that some of our design goals were
in conflict. It was clear that concept-based overloading was an
important language feature, but it seemed to conflict with being able
to separately compile templates. Even worse, there was tension between
concept-based overloading and modular type checking.

\begin{figure}[tbp]
  \centering
\begin{lstlisting}
template <InputIterator Iter1, OutputIterator Iter2>
Iter2 copy(Iter1 first, Iter1 last, Iter2 result);

template <RandomAccessIterator Iter1, OutputIterator Iter2>
Iter2 copy(Iter1 first, Iter1 last, Iter2 result);

template <InputIterator InIter1, InputIterator InIter2, OutputIterator OutIter>
OutIter merge(InIter1 first1, InIter1 last1, InIter2 first2, InIter2 last2, OutIter result) {
  ...
  return copy(first2, last2, copy(first1, last1, result));
}  
\end{lstlisting}
  \caption{An example that demonstrates late versus early-bound overloading.}
  \label{fig:late-early-overloading}
\end{figure}

Consider the excerpt from the STL \code{merge} function together with
two overloads of the \code{copy} function template in
Figure~\ref{fig:late-early-overloading}. To separately compile
\code{merge}, the \cpp{} compiler needs to decide which overload
of \code{copy} should be called without knowing the actual iterator
types. If the \cpp{} compiler performed overload resolution only
based on the type constraints for \code{merge}, it would resolve
to the overload of copy for Input Iterator. However, consider what
happens when using \code{merge} with the iterators from a vector,
which model Random Access Iterator. With the resolution already
decided, \code{merge} would call the slow version of \code{copy}.

Suppose one forsakes separate compilation in favor of a design that
allows late binding (after instantiation) for concept-based
overloading. The next question is whether such a design can support
modular type checking. The answer to this question is not
straightforward, and turns out to be analogous to the problem of
modular type checking in languages with
multi-methods~\cite{Chambers:1995uq,Millstein:1999kx,Clifton:2006vn,Millstein:2009ys}.
It is also analogous to the problem of dealing with overlapping
instance declarations in Haskell~\cite{jones97type}. It is possible to
achieve modular type checking, but the resulting type systems are
complicated.

Facing this fork in the design space, we split in two directions: I
investigated a design with separate compilation and early-bound
overloads and J\"arvi and Willcock explored a design with late-bound
overloads but not separate compilation.  My investigation led to the
design of a calculus for generics, named
$\mathrm{F}^{\mathcal{G}}$~\cite{Siek:2005mf,siek04:_fg1} and a
prototype language named
$\mathcal{G}$~\cite{Siek:2005lr,siek05:_g_stl,Siek:2008rr}. The
exploration by J\"arvi and Willcock did not immediately lead to a
concrete design, but their work later became the foundation for the
Indiana proposal for concepts in \cpp{}, which we discuss in
Section~\ref{sec:indiana-proposal}.

\input{fg}

\subsection{A Prototype Language for Generic Programming}
\label{sec:g}

To gain some practical experience in using this language design, I
developed a prototype language named
$\mathcal{G}$~\cite{Siek:2005lr,siek05:_g_stl,Siek:2008rr}. Then, as a
case study, I ported a large portion of the STL to $\mathcal{G}$. In
addition to the features in $\mathrm{F}^{\mathcal{G}}$, $\mathcal{G}$
included template argument deduction, function overloading, concept
refinement, parameterized models, a basic module system, and simple
classes.  The overall experience of porting the STL to $\mathcal{G}$
was pleasant: short error messages and short compile times.

As mentioned earlier, in deciding to support separation compilation,
the design for $\mathcal{G}$ gave up late-bound concept-based
overloading. However, $\mathcal{G}$ does provide early-bound
concept-based overloading, and in the process of porting the STL to
$\mathcal{G}$, I discovered workarounds for mimicking late-bound
overloads using early-bound overloads.

Recall the \code{merge} function from
Figure~\ref{fig:late-early-overloading}. When ported to $\mathcal{G}$,
the calls to \code{copy} always resolve to the slower version for
Input Iterator, even when \code{merge} is used on a sequence that
models Random Access Iterator.
Figure~\ref{fig:mimick-late-overloading} shows the workaround for
mimicking late-bound overloading. The main idea is to create a new
concept for the overloaded function, in this case for \code{copy}.
Here we name the concept \code{CopyRange}. Instead of calling
\code{copy} directly, \code{merge} adds \code{CopyRange} to its
constraints (in $\mathcal{G}$, constraints go in the \code{where}
clause), and then calls the \code{copy_range} operation instead of
\code{copy}.

With this change, the choice of which model of \code{CopyRange} is
used, and therefore which \code{copy}, is made at the point where
\code{merge} is instantiated, where the concrete iterator is known.
However, it would place a burden on users of \code{merge} if they had
to create extra model definitions for \code{CopyRange}. To solve this
problem, we create two parameterized model definitions for
\code{CopyRange}, one for models of Input Iterator and one for models
of Random Access Iterator. With these two parameterized models, the
user of \code{merge} does not need to do any extra work.

\begin{figure}[tbp]
  \centering
\begin{lstlisting}[morekeywords={model,fun,where}]
concept CopyRange<I1,I2> {
  fun copy_range(I1,I1,I2) $\mathsf{-}$> I2;
};

model <Iter1,Iter2> where { InputIterator<Iter1>, ... }
CopyRange<Iter1,Iter2> {
  fun copy_range(Iter1 first, Iter1 last, Iter2 result) $\mathsf{-}$> Iter2
    { return copy(first, last, result); }
};

model <Iter1,Iter2> where { RandomAccessIterator<Iter1>, ... }
CopyRange<Iter1,Iter2> {
  fun copy_range(Iter1 first, Iter1 last, Iter2 result) $\mathsf{-}$> Iter2
    { return copy(first, last, result); }
};

fun merge<Iter1,Iter2,Iter3>
where { ..., CopyRange<Iter2,Iter3>, CopyRange<Iter1,Iter3> }
(Iter1 first1, Iter1 last1, Iter2 first2, Iter2 last2, Iter3 result) $\mathsf{-}$> Iter3 {
  ... return copy_range(first2, last2, copy_range(first1, last1, result));
}
\end{lstlisting}
  \caption{Workaround for mimicking late-bound overloading.}
  \label{fig:mimick-late-overloading}
\end{figure}

The attentive reader may wonder how \code{merge}, using the above
workaround, can be separately compiled. The reason is that
$\mathcal{G}$ uses the same dictionary-passing implementation model as
$\mathrm{F}^{\mathcal{G}}$. Thus, $\mathcal{G}$ generates just one
version of \code{merge} that takes a dictionary for
\code{CopyRange}, which includes a pointer to the appropriate
\code{copy_range} function. The cost of separate compilation, and this
dictionary-passing model, is an extra level of indirection and
therefore some run-time overhead. However, the $\mathcal{G}$ compiler
is free to optimize calls to \code{merge}, through function
specialization and constant-propagation through the dictionaries, to
remove the run-time overhead in situations where performance outweighs
separate compilation.  \citet{Jones:1995uq} describes this
optimization in the context of Haskell type classes.

\section{The Indiana Proposal}
\label{sec:indiana-proposal}

After the Kona meeting in 2003, two \cpp{} standards meetings (one
year) elapsed without an updated proposal from the Texas A\&M team.
In the Fall of 2004, the Indiana team began to worry about the
progress on concepts with respect to getting an extension into
\cppox. Around the same time, Douglas Gregor arrived at Indiana
University to do a post-doc with Andrew Lumsdaine, and Douglas was
excited to join in the design effort and to implement concepts in
the GNU \cpp{} compiler. The prototype would be named ConceptGCC.
The Indiana team began in earnest to develop a proposal for \cppox{}
in the Fall of 2004.  The design differed from $\mathcal{G}$ in
several respects, which we discuss in the next few paragraphs.

Achieving separate compilation for \cpp{} was a non-starter because we
had to maintain backwards compatibility, and existing features of
\cpp{} such as user-defined template specializations interfere with
separate compilation (as discussed in
Section~\ref{sec:template-spec-separate-comp}). With that in mind, it
made sense to use the late-bound approach to concept-based overload
resolution~\cite{jaakko06:_algo_spec}.

In general, performance is a high priority for \cpp{} programmers, so
we needed an implementation that would yield zero run-time
overhead. The dictionary-passing implementation used in $\mathcal{G}$
comes with roughly the same overhead as virtual method dispatch.  To
improve upon this, we devised an approach that instead resolves all uses of
concept operations at compile
time~\cite{gregor05:implementing_concepts}.

The third difference between the Indiana proposal for \cpp{} and
$\mathcal{G}$ was that models would not be lexically
scoped but instead they would reside in the same namespace as their
concept. This design choice was made primarily to simplify the
implementation but it meant that model definitions would not be
modular. 

Compared to the Texas proposal, there were also several differences.
Instead of usage patterns to specify concept operations, the Indiana
proposal went with pseudo-signatures, a design that the Texas team had
considered but discarded~\cite{stroustrup_concept_checking03}.  The
idea with pseudo-signatures is to relax the rules for matching up
functions to concept operations. Consider the example in
Figure~\ref{fig:pseudo-sig} in which class \code{A} models concept
\code{C} (written in the syntax of the Indiana proposal).  The concept
requires a function \code{f} taking two parameters of type \code{A}
and \code{B} and has return type \code{A}. (In this case, \code{A} is
substituted for \code{T}.)  However, the only function named \code{f}
in the scope of the model definition has both parameters at type
\code{A} and a return type of \code{B}. With pseudo-signatures, this
function \code{f} satisfies the requirement because \code{B} is
implicitly convertible to \code{A}.

The strength of pseudo-signatures is that it provides flexibility for
the client of a template while at the same time providing precision for
the implementor of a template.
The pseudo-signature approach was straightforward to implement.  On
the client side, finding implementations that match a pseudo-signature
can be accomplished with the normal algorithm for function overload
resolution. On the implementation side of a template, a
pseudo-signature is treated just like a normal signature for the
purposes of type checking. That is, what you see is what you get.

\begin{figure}[tbp]
  \centering
\begin{lstlisting}[morekeywords={model,typeid}]
class A {};
class B : public A {};

template<typeid T>
concept C {
  A f(T, B);
};

B f(A, A) { return B(); }

model C<A> { };
// This model is OK because B is convertible to A.
\end{lstlisting}
  \caption{Example of a pseudo-signature.}
  \label{fig:pseudo-sig}
\end{figure}

The Indiana proposal also differed from the Texas proposal in that it
did not support implicit modeling, but instead relied on model
definitions to establish the modeling relationship between a class and
a concept. We did not want to expose programmers to the kind of
accidental run-time errors discussed in Section~\ref{sec:kona}.
With respect to operators for combining concepts, the Indiana proposal
only included conjunction.

\begin{figure}[tbp]
  \centering
\begin{lstlisting}[morekeywords={where,require}]
template<typeid Iter> 
concept InputIterator : EqualityComparable<Iter>, 
    CopyConstructible<Iter>, Assignable<Iter> { 
  typename value_type;
  typename reference; 
  require Convertible<reference, value_type>; 
  Iter& operator++(Iter&); 
  reference operator*(Iter);
}; 

template<typeid InIter, typeid OutIter> 
where { InputIterator<InIter>, OutputIterator<OutIter>, 
         InIter::value_type == OutIter::value_type } 
OutIter copy(InIter a, InIter b, OutIter out) { 
  while (a != b) *out++ = *a++; 
  return out; 
} 
\end{lstlisting}
  \caption{Example of a concept and constrained template in the Indiana proposal. }
  \label{fig:indiana-eg}
\end{figure}

Figure~\ref{fig:indiana-eg} shows an example of a concept definition
and constrained template from the Indiana proposal.

The Indiana proposal was completed and submitted as document number
N1758 to the \cpp{} standards committee in January of
2005~\cite{siek05:concepts_c++0x}.

\section{The Texas Rebuttal}
\label{sec:texas-rebuttal}

\citet{stroustrup05:concept_design_rev_1} submitted a revised proposal
in April of 2005 and submitted a paper describing their design to POPL
2006~\cite{Dos-Reis:2006zr,Dos-Reis:2005uq}. The proposal moved closer
to the Indiana proposal in several respects, but there were
significant differences.  The revised proposal included \code{where}
clauses to enable the use of multi-type concepts and added
associated types.
The revised proposal also added model declarations, under the name
\code{static_assert}. However, unlike the Indiana proposal, model
declarations were optional, so the Texas proposal retained implicit
modeling.
The Texas proposal kept usage patterns for specifying concept
operations and an associated technical report defined an algorithm for
converting usage patterns into type signatures~\cite{Dos-Reis:2005uq}.
The proposal also retained the \code{||} and \code{!} operators for
combining concepts.


To address the concern regarding accidental run-time errors, discussed
in Section~\ref{sec:kona}, \citet{stroustrup05:concept_design_rev_1}
proposed using negative model declarations, that is, declarations that
a type does not implement a concept. For the example in
Section~\ref{sec:kona}, they suggested adding a declaration stating
that \code{istream_iterator} does not model \code{ForwardIterator}.

At the Mont Tremblant \cpp{} standards meeting in October 2005, there
was considerable disagreement regarding the design of concepts. The
Indiana team was not in favor of negative model declarations because
the number of properties that a type does not satisfy is open ended
and much larger than the number of properties that a type does
satisfies.  Thus, negative model declarations did not seem like an
economical approach to solving this problem. 

With respect to usage patterns and the \code{||} and \code{!}
operators for combining concepts, there was no publicly available
compiler implementation, so doubt remained regarding the usability and
efficiency of these design choices.

\section{The Compromise at Adobe}
\label{sec:adobe}

Alexander Stepanov invited the Texas and Indiana teams to a meeting at
Adobe Systems Inc. in San Jose. The goal of the meeting was to resolve
the outstanding differences between the two proposals, and indeed, the
teams were able to agree on a compromise design.  The high points of
the compromise were as follows. The joint design would include
\code{||} and \code{!} operators for combining concepts, but it would
use pseudo-signatures instead of usage patterns.  With respect to
model declarations, the compromise was to have two kinds of concepts.
The default kind of concept would require explicit model declarations
whereas a concept that started with the keyword \code{auto} could be
modeled implicitly.  The Indiana team was receptive to this compromise
because it balanced convenience and safety.

Over the next few months, the Texas and Indiana teams worked together
to document the compromise design and published the results at OOPSLA
2006~\cite{gregor06:_concepts}. One of the syntactic changes that
happened during this time, thanks to research by Beman Dawes, was that
the keyword \code{model} was replaced by \code{concept_map} to
minimize the number of existing \cpp{} programs that would break when
upgraded to the new version of \cpp{}.

To avoid ``too many cooks in the kitchen'', Stroustrup and Gregor
teamed up to write the compromise proposal to the \cpp{} standards
committee, which resulted in the document N2042~\cite{Gregor:2006ly}
in June of 2006 and the revision N2081 in September. The proposal was
well received by the \cpp{} standards committee.



\section{Proposed Wording and Acceptance}
\label{sec:acceptance}

In 2007, work began in earnest to draft the wording that would go into
the \cpp{} standard. The dialect of English used in the \cpp{}
standard is lovingly referred to as ``standardese''. Unfortunately,
writing standardese is a slow process. Gregor and Stroustrup wrote the
initial draft and several revisions, producing the documents N2193,
N2307, and N2398. In the Fall of 2007, Widman and I began to pitch
in with the writing, and helped produce N2421, N2501, and N2520.

During this time, the design evolved in a few respects. The
\code{where} keyword was changed to \code{requires} because 
\code{requires} would cause fewer existing \cpp{}
programs to break.
Also, the \code{||} constraints were removed for lack of
implementation experience.

The initial implementation approach for concepts required the
generation of forwarding functions within each
model~\cite{gregor05:implementing_concepts}. While these function
calls can be statically resolved (and usually inlined), there was
still some run-time overhead with ConceptGCC. Also, some optimizations
such as copy-elision could not be applied with the forwarding functions
in the way. To address these issues, Gregor devised an alternative
compilation approach that did not rely on forwarding functions but
instead inlines the body of the forwarding functions into
the template~\cite{Gregor:2008fk}. Unfortunately, this approach opened
up another hole in the modularity of the type system: there could be
ambiguities after instantiation that cause type errors.

Meanwhile, I revisited the decision that model declaration would live
in the same namespace as their concept, proposing to allow model
declarations to appear in any namespace and to use the normal
(lexical) name lookup rules for concept maps~\cite{Siek:2006qf}. This
proposal was referred to as ``scoped concept maps''. In 2007, scoped
concept maps were approved and proposed wording for the standard was
drafted by Widman and I~\cite{Siek:2007uq}.

In addition to the work on the concepts feature itself, there was
considerable work to update the \cpp{} standard library to use
concepts. Gregor, Halpern, Lumsdaine, Marcus, Witt, and I with the
help of many others, drafted the changes to the \cpp{} standard,
producing N2500, N2502, N2677, and their revisions.

At the September 2008 meeting of the \cpp{} standards committee in San
Francisco, the concepts proposal was voted into the working draft for
\cppox! In the next few months, the editor of the working draft, Pete
Becker, began merging the proposed wording for concepts and the
updates to the standard library into the \cpp{} standard. At this
point, we believed it would be smooth sailing, with the remaining work
on the order of fixing typos and resolving ambiguities in the wording.
We hoped that the entire working draft would be finalized in a year or
so and then approved as the new ANSI and ISO standard for \cpp{}.

\section{The Removal of Concepts}
\label{sec:removal}

In the months following the San Francisco meeting, heated discussions
occurred on the \cpp{} standards committee mailing list. There were two
threads of discussion. The first thread was kicked off by Howard
Hinnant, with the email titled ``Are concepts required of Joe
Coder?''. The question was whether a programmer using the standard
library would need to be aware of concepts. 
My simple answer is yes. Concepts inform the programmer regarding
which types can be used with which templates.  In fact, before the
addition of the concepts feature to \cpp{}, programmers needed to be
aware of concepts (in the form of documentation) for the same reason.

However, there was a concern that the addition of concepts would
create a learning curve that would be too great for the average \cpp{}
programmer. In particular, there was concern that it would be too
burdensome for programmers to write lots of concept maps.
So an important question is: in which situations does a programmer
need to write concepts maps?

The most common place for concept maps is immediately after a class
definition. The author of the class knows that the class models
various concepts and therefore documents these facts for users of the
class, including the \cpp{} compiler.  This use of concept maps is
analogous to a class inheriting from an abstract base class (that is,
an interface). Millions of programmers have learned object-oriented
languages and inheritance, so learning to use concept maps cannot be
too great of a hurdle.  Furthermore, many concepts would be
\lstinline{auto} concepts, for which concept maps are not required.

A less common place for concept maps is when a programmer wants to use
one library with another library, but the library authors did not
intended the libraries to be used together. In such cases a programmer
can use concept maps to retroactively specify that a class implements
a concept. 
Retroactive inheritance has long been desired in object-oriented
languages. The \emph{external polymorphism} design pattern provides a
workaround~\cite{Gamma:1995vn} for languages without retroactive
inheritance and there have been many language extensions that provide
retroactive inheritance, such as
signatures~\cite{Baumgartner:1995fk,Laufer:2000uq} and
aspects~\cite{Kiczales:1997lh}.
So this use of concept maps is a nice advance compared to traditional
forms of inheritance.


In a reaction to the thread ``Are concepts required of Joe Coder?''
and to move closer to his original design, Stroustrup proposed to
remove explicit concepts (concepts that require concept maps) and
replace them with \emph{explicit refinement}~\cite{Stroustrup:2009ys}.
However, the semantics of explicit refinement was not clear, so it
was very difficult for committee members to evaluate the proposal.

The second thread of discussion concerned the state of implementations
of the concepts feature~\cite{Tasker:2009zr}. Although Gregor had
implemented a prototype, ConceptGCC, the concept specification had
moved beyond the prototype with changes and additions. Also, there
were problems with the prototype: bugs and slow compile times, that
made it difficult to use ConceptGCC with large generic libraries.
(Gregor is not to blame in this regard, as implementing concepts
inside the Gnu C compiler was a heroic task.)  The slow compile times
also worried many committee members, even though it was an engineering
issue and not a theoretical limit that was causing the slow down.  An
algorithm for fast type checking had been demonstrated in the
$\mathcal{G}$ prototype~\cite{Siek:2005mf,Siek:2005lr}.

At the Frankfurt meeting in July 2009, the \cpp{} standard committee
voted with the following options.
\begin{enumerate}\addtolength{\itemsep}{-0.5\baselineskip}
\item Continue with the current specification of concepts.
\item Remove explicit concepts and add explicit refinement.
\item Remove concepts from the working draft of the \cpp{} standard.
\end{enumerate}
Most of the committee members felt it was too late for major changes,
and without Stroustrup supporting the status quo, the overwhelming
majority voted to remove concepts. Needless to say, everyone who had
been involved in the development of concepts was deeply disappointed.

\section{Conclusion}
\label{sec:conclusion}

So where do concepts go from here? The next round of \cpp{}
standardization will most likely be in five years.  However, the more
important question is whether \cpp{} programmers, and the \cpp{}
committee in particular, will be able to gain experience using
concepts so that they can better evaluate the tradeoffs regarding
different designs. So there is great need for an implementation of
concepts. However, for the next few years, most commercial \cpp{}
compiler vendors will be focused on implementing the features that
made it into \cppox{}. Also, implementing concepts is an
expensive endeavor, and without consensus in the \cpp{} committee on
the design, such an expense is risky from a business point of view.

There is some hope that Douglas Gregor, heading up the \cpp{} compiler
team at Apple, Inc., together with Andrew Lumsdaine's research group
at Indiana University, will be able to implement concepts within the
Clang LLVM-based compiler.
On another front, I am leading a group at the University of Colorado
to add concepts to the Chapel programming
language~\cite{Chamberlain:2005fd} with funding from the U.S.
Department of Defense.  While Chapel is somewhat different from
\cpp{}, this may give some programmers more exposure to concepts and
generate feedback regarding the design.

\section*{Acknowledgments}

Thank you to Jeremy Gibbons for organizing the Spring School on
Generic and Indexed Programming and for inviting me to give a lecture
there on the \CPPOX{} concept effort. Thank you Oleg Kiselyov for your
many kind suggestions for improvement for this article. Thanks to
Bjarne Stroustrup for his corrections and thanks to the programming
languages group at the University of Colorado for your helpful
feedback and proof reading. The writing of this article was supported
by the NSF in the form of a CAREER award, CCF 0846121.

{\footnotesize
\bibliographystyle{abbrvnat}
\bibliography{all}
}

\end{document}

%% file: defs.tex
\lstdefinestyle{basic}{showstringspaces=false,columns=fullflexible,language=ML,
escapechar=$,xleftmargin=1pc,%
mathescape=true,%
basicstyle=\small\sffamily,%
commentstyle=\mdseries,%
morekeywords={object,with,fun,new,end,int,bool,string},%
deletekeywords={error,apply,values,and,equal,eq},%
moredelim=**[is][\color{white}]{~}{~},%
}
\lstdefinestyle{cpp}{showstringspaces=false,columns=fullflexible,language=C++,
escapechar=$,xleftmargin=1pc,%
mathescape=true,%
basicstyle=\small\sffamily,%
morekeywords={concept},%
commentstyle=\mdseries,%
moredelim=**[is][\color{white}]{~}{~},%
}

\newcommand{\LT}[0]{\texttt{<}}
\newcommand{\GT}[0]{\texttt{>}}

\newcommand{\concepta}[5]{\mathbf{\mathsf{concept}}\; #1\LT #2 \GT \{ #3;\;#4;\; #5\}}

\newcommand{\modela}[4]{\mathbf{\mathsf{model}}\; #1\LT#2\GT \; \{ #4;\; #3 \}}

\newtheorem{theorem}{Theorem}
\newtheorem{lemma}{Lemma}

%% file: comparison_chart.tex
\newcolumntype{C}{>{\hfill\arraybackslash}p{.4in}<{\hfill\hfill}}

\newcommand{\featY}{\CIRCLE} 
\newcommand{\featP}{\begin{sideways}\LEFTcircle\end{sideways}} 
\newcommand{\featN}{\Circle} 

\newcommand{\featA}{\hspace{4.25pt}\featY$^{*}$} 
\newcommand{\featB}{\hspace{4.25pt}\featY$^{\dagger}$} 

\begin{table}[t]
\center\footnotesize
\begin{tabular}{rcccccc}
\hline \hline
{} &  C++ & SML & Haskell & Eiffel & Java & C\# \\
\hline
Multi-type concepts    & -      & \featY & \featA & \featN & \featN & \featN  \\
Multiple constraints   & -      & \featP & \featY & \featN & \featY & \featY  \\
Associated type access & \featY & \featY & \featP & \featP & \featP & \featP  \\
Retroactive modeling   & -      & \featY & \featY & \featN & \featN & \featN  \\
Type aliases           & \featY & \featY & \featY & \featN & \featN & \featN  \\
Separate compilation   & \featN & \featY & \featY & \featY & \featY & \featY  \\
Template arg. deduction & \featY & \featN & \featY & \featN & \featY & \featY \\ 
Concise syntax & \featY & \featP & \featY & \featN & \featP & \featN \\
\hline \hline
\end{tabular} \\
\raggedright {\small
$^{*}$Using the multi-parameter type class extension to Haskell.
}
\caption{Results from the study of language support
for generic programming. A black circle indicates full support, a white circle indicates poor
support, and a half-filled circle indicates partial support.
The rating of ``-'' in the \protect\cpp{} column indicates that
\protect\cpp{} does not explicitly support the feature, but one can
still program
as if the feature were supported due to the permissiveness of
\protect\cpp{} templates.\label{tab:study}}
\end{table}

%% file: fg.tex
\subsection{A Calculus for Generic Programming}

Here we present a streamlined version of $\mathrm{F}^{\mathcal{G}}$ to
give the reader a feel for the design work that went on at Indiana.
The purpose of a calculus like $\mathrm{F}^{\mathcal{G}}$ is not to
define a language that is for programmers to use, but it is instead a
mathematical tool to help a language designer better understand the
impact of different design decisions. In particular,
$\mathrm{F}^{\mathcal{G}}$ was meant to check whether my language
design provided modular type checking. In addition to informing the
design for concepts, $\mathrm{F}^{\mathcal{G}}$ inspired the
\emph{implicits} feature of the Scala
language~\cite{scala-overview-tech-report}.

The inspiration for $\mathrm{F}^{\mathcal{G}}$ came primarily from
Haskell and Standard ML. The $\mathrm{F}^{\mathcal{G}}$ language
borrows type classes and instance declarations from Haskell, with some
modification, to form concepts and models. The main differences are
that 1) $\mathrm{F}^{\mathcal{G}}$ does not infer constraints but
instead allows an operation name to appear in more than one concept,
and 2) model declarations are lexically scoped in
$\mathrm{F}^{\mathcal{G}}$ instead of being essentially global. With
respect to Standard ML, $\mathrm{F}^{\mathcal{G}}$ borrows abstract
types and type sharing from ML signatures to provide support for
associated types.

Figure~\ref{fig:fg-syntax} defines the syntax of
$\mathrm{F}^{\mathcal{G}}$. As the name implies,
$\mathrm{F}^{\mathcal{G}}$ is an extension to System
F~\cite{GIRARD72,REYNOLDS74C}, which we reviewed in
Section~\ref{sec:intro}. $\mathrm{F}^{\mathcal{G}}$ is also inspired
by the qualified types of \citet{jones94:_qualif_types}. For
$\mathrm{F}^{\mathcal{G}}$, we retain the explicit instantiation
syntax of System F ($e[\tau]$) to avoid complexity in well-understood
areas of the design.  The language $\mathcal{G}$ (presented in
Section~\ref{sec:g}), however, does provides implicit instantiation
(template argument deduction).  The overbar notation $\overline{a}$
indicates a list of items in the syntactic category $a$. For example,
$\overline{\tau}$ is a list of types: $\tau_1,\tau_2,\ldots,\tau_n$.

$\mathrm{F}^{\mathcal{G}}$ adds several syntactic forms to System F:
$C \Rightarrow e$ is a constrained expression, with the syntactic category $C$
for expressing constraints.  A constraint is either a concept
constraint of the form $c\LT \overline{\tau} \GT$ or a same-type
constraint of the form $\tau = \tau$.  The expression $\pi$ is a term
identifier, which is a variable possibly prefixed by a sequence of
model identifiers. The purpose of such expressions is to refer to
members of a model.  The expression
$(\concepta{c}{\overline{\alpha}}{\overline{\beta}}{\overline{C}}{\overline{x:T}}
\;\mathsf{in}\; e)$ defines a concept named $c$ for use in $e$. The
$\beta$'s are requirements for associated types, $\overline{C}$ are
nested constraints, and $\overline{x : T}$ are the required members of
the concept. The expression
$(\modela{c}{\overline{T}}{\overline{x = e}}{\overline{\beta =
    T}}\;\mathsf{in}\; e)$ establishes that the types $\overline{T}$
together model the concept $c$. (The common case is for there to
be only one modeling type.) The expression $(\mathsf{type} \; \alpha = \tau
\;\mathsf{in}\; e)$ aliases the type $\tau$ to the variable $\alpha$. (In
some sense, type aliases are not necessary, because the programmer
could manually substitute $\tau$ for $\alpha$ in $e$.  But type alias are
important from a software engineering point of view because, in
generic programming, type expressions can grow quite large, so typing
$\tau$ many times inside $e$, and maintaining the resulting code,
becomes rather burdensome~\cite{Garcia:2003gn,Garcia:2007fk}.)

\begin{figure}[tbp]
  \centering
\[
\begin{array}{llcl}
\text{concept names} & c  \\
\text{integers} & n \\
\text{types} &  \tau  &::=& \mathsf{int} \mid \tau -> \tau \mid \forall \alpha. \; \tau \mid C \Rightarrow \tau \mid \Pi \\
\text{model identifier} & m & ::= & c \LT\overline{\tau}\GT \\
\text{type identifier} & \Pi & ::= & \alpha \mid m . \Pi \\
\text{constraints} & C &::=& m \mid \tau = \tau  \\
\text{expressions} &  e &::=& n \mid e\; e \mid \lambda y\!:\!\tau.\; e \mid \Lambda \alpha.\; e \mid e[\tau] \\
     & &\;|& C \Rightarrow e \mid \pi \\
     & &\;|& \concepta{c}{\overline{\alpha}}{\overline{\beta}}{\overline{C}}{\overline{x:\tau}} \;\mathbf{\mathsf{in}}\; e\\
    &  &\;|& \modela{c}{\overline{\tau}}{\overline{x = e}}{\overline{\beta = \tau}}\;\mathbf{\mathsf{in}}\; e \\
    & & \;| & \mathbf{\mathsf{type}}\; \alpha = \tau \;\mathbf{\mathsf{in}}\; e \\
\text{term identifier} & \pi &::= &  x \mid m . \pi \\
\text{environments} &  \Gamma & ::= &  \Gamma,x:\tau \mid \Gamma,\alpha \mid \Gamma, C \\
  & & \;| & \Gamma, \concepta{c}{\overline{\alpha}}{\overline{\beta}}{ \overline{C}}{\overline{x : \tau}}
\end{array}
\]
  
  \caption{Syntax of $\mathrm{F}^{\mathcal{G}}$.}
  \label{fig:fg-syntax}
\end{figure}


The type system for $\mathrm{F}^{\mathcal{G}}$ is defined in
Figure~\ref{fig:fg-type}. As can be seen in the rule for function
application, type equality is not simply syntactic equality in
$\mathrm{F}^{\mathcal{G}}$; the addition of same-type constraints
means that the type system needs to perform equational reasoning over
types. The judgment $\Gamma \vdash \tau = \tau$ gives the declarative rules for
type equality. These rules can be implemented efficiently using a
congruence closure
algorithm~\cite{nelson80:_fast_cong_clos,Downey:1980fk}.  Otherwise,
the typing rules for function and generics is the same as System F.

The introduction (creation) of an expression of type $C \Rightarrow
\tau$ is explicit whereas the elimination (use) of such an expression
is implicit.  This design is intended to make it convenient to use
generic libraries without adding too much complexity to the type
system. (We also want to avoid overburdening implementors of the
language). To type check a constrained expression $C \Rightarrow e$,
we type check $e$ with $C$ added as an assumption in the environment.
If $C$ is a concept constraint, then we also add all constraints that
are nested inside the concept into the environment, using the notation
$\flat(C)_\Gamma$ defined as follows:
\begin{align*}
  \flat(\tau_1 = \tau_2)_\Gamma &= (\tau_1 = \tau_2) \\
  \flat(c \LT \overline{\tau} \GT)_\Gamma  &= 
    c \LT \overline{\tau} \GT, [\overline{\alpha} {:=} \overline{\tau}]\overline{\flat(C)_\Gamma }
 & \text{if } \concepta{c}{\overline{\alpha}}{\overline{\beta}}{\overline{C}}{\overline{x:T}} \in \Gamma
\end{align*}
To type check the elimination (use) of an expression of type $C \Rightarrow
\tau$, we check that that the constraint $C$ can be satisfied in the
current environment, for which we use the judgment $\Gamma \vdash C$.  A
concept constraint $c \LT\overline{\tau}\GT$ is satisfied if there is a
model for that model identifier in the environment.  A same-type
constraint $\tau_1 = \tau_2$ is satisfied if the types are equal in
the current environment.

Note how the typing rules for introducing and eliminating constrained
expression are nearly mirror images of each other. The introduction
rule makes some assumptions and the elimination rule discharges an
assumption. This kind of mirror imaging is required to create a type
system that ensures type safety. With that in mind, it is worth noting
a discrepancy here: the elimination rule only discharges one
assumption whereas the introduction rule adds many assumptions to the
environment. The extra assumptions come from nested constraints, and
if you look at the typing rule for models you see the check for the
nested constraints there.

Type checking a concept definition expression is straightforward, we
add the concept to the environment and proceed to check the rest of
the program. Type checking a model definition expression, of the
following form, deserves some explanation.
\[
\modela{c}{\overline{\tau_1}}{\overline{x = e}}{\overline{\beta = \tau_3}}\;\mathsf{in}\; e
\]
First, there must be a concept named $c$ in the environment.  Second,
the nested constraints $C$ within $c$ must be satisfied in the current
environment. Third, the expressions $\overline{e}$ must be of the
right types, as specified by the concept. Finally, we type check the
rest of the program $e$ in an environment where we have added the model
$c\LT\overline{\tau}\GT$ as an assumption, as well as the type
equalities for its choices regarding the associated types.

In the definition of the $\mathit{lookup}$ function, the notation
$\Gamma|_c$ means the environment obtained by removing everything but
the concept and same-type constraints from $\Gamma$.

\begin{figure}[tbp]
  \small
\fbox{$\Gamma \vdash e : \tau$}
\begin{gather*}
  \inference{}{\Gamma \vdash n : \mathsf{int}}\quad
  \inference{\Gamma,x:\tau  \vdash e : \tau'}{\Gamma  \vdash \lambda x:\tau. e : \tau \to \tau'} \quad
  \inference{\Gamma \vdash e_1 : \tau_1 \to \tau_2  & \Gamma \vdash e_2 : \tau_3 \\ \Gamma \vdash \tau_1 = \tau_3}
      {\Gamma \vdash e_1 \; e_2 : \tau_2}\\[2ex]
  \inference{\Gamma, \alpha \vdash e : \tau}
     {\Gamma \vdash \Lambda \alpha.\; e : \forall \alpha.\; \tau} \quad
  \inference{\Gamma \vdash e : \forall \alpha.\; \tau}{\Gamma \vdash e[\tau'] : [\alpha:=\tau']\tau} \\[2ex]
  \inference{\Gamma, \flat(C)_\Gamma \vdash e : \tau}
            {\Gamma \vdash C \Rightarrow e : C \Rightarrow \tau}
\quad
  \inference{\Gamma \vdash e : C \Rightarrow \tau & \Gamma \vdash C}
            {\Gamma \vdash e : \tau}
   \\[1ex]
  \inference{\Gamma, \alpha = \tau' \vdash e : \tau}
      {\Gamma \vdash \mathsf{type}\;\alpha = \tau' \;\mathsf{in}\; e : \tau}
  \quad
 \inference{
    \mathit{lookup}(\pi,\Gamma) = \tau
 }{
   \Gamma \vdash \pi : \tau
 }
\\[2ex]
  \inference{\Gamma, \concepta{c}{\overline{\alpha}}{\overline{\beta}}{\overline{C}}{\overline{x:S}} \vdash e : T}{\Gamma \vdash \concepta{c}{\overline{\alpha}}{\overline{\beta}}{\overline{C}}{\overline{x:S}} \;\mathsf{in}\; e : T} 
\\[2ex]
  \inference{
    \concepta{c}{\overline{\alpha}}{\overline{\beta}}{\overline{C}}{\overline{x:\tau_2}} \in \Gamma &
    \Gamma,\overline{\alpha=\tau_1},\overline{\beta = \tau_3} \vdash \overline{C} \\
    \Gamma,\overline{\alpha=\tau_1},\overline{\beta = \tau_3} \vdash \overline{e} : \overline{\tau_2} &
    \Gamma, c \LT \overline{\tau_1} \GT, \overline{c \LT \overline{\tau_1} \GT.\beta = \tau_3} \vdash e : \tau_4}
  {\Gamma \vdash \modela{c}{\overline{\tau_1}}{\overline{x = e}}{\overline{\beta = \tau_3}}\;\mathsf{in}\; e : \tau_4}
\end{gather*}
\boxed{\Gamma \vdash C}\qquad
  \inference{c \LT \overline{\tau'} \GT \in \Gamma & \Gamma \vdash \overline{\tau = \tau'}}
     {\Gamma \vdash c \LT \overline{\tau} \GT} \quad
  \inference{\Gamma \vdash \tau_1 = \tau_2}{\Gamma \vdash (\tau_1 = \tau_2)}
 \\[2ex]
\fbox{$\Gamma \vdash \tau = \tau$} $\qquad$ The congruence closure (with respect
to $\to,\forall,\Rightarrow$) of the following:
\begin{gather*}
\inference{}{\Gamma \vdash \alpha = \alpha} \qquad
\inference{\tau_1 = \tau_2 \in \Gamma}{\Gamma \vdash \tau_1 = \tau_2} \qquad
\inference{\Gamma \vdash \overline{\tau_1} = \overline{\tau_2} & \Gamma \vdash \Pi_1 = \Pi_2}
 {\Gamma \vdash c\LT\overline{\tau_1}\GT.\Pi_1  = c\LT\overline{\tau_2}\GT.\Pi_2}
\end{gather*}
\fbox{$\mathit{lookup}(\pi,\Gamma) = \tau$}
\begin{align*}
  \mathit{lookup}(x, \Gamma) & =  \tau   &  \text{if } x : \tau \in \Gamma \\ 
  \mathit{lookup}(c \LT \overline{\tau} \GT . \pi, \Gamma) & = 
   \mathit{lookup}(\pi, (\Gamma|_{c},\overline{\alpha = \tau},\overline{C},\overline{x : \tau'})) \\
  & \text{if } \Gamma \vdash c \LT \overline{\tau} \GT 
   \text{ and } \concepta{c}{\overline{\alpha}}{\overline{\beta}}{\overline{C}}{\overline{x:\tau'}} \in \Gamma   
\end{align*}
  \caption{The type system of $\mathrm{F}^{\mathcal{G}}$.}
  \label{fig:fg-type}
\end{figure}

Although $\mathrm{F}^{\mathcal{G}}$ is a small calculus, with the
addition of a few standard features such as lists, integers, and
fix~\cite{Pierce:2002hj}, it can express simple generic algorithms.
Figure~\ref{fig:fg-example} shows the definition of a generic
\code{foldl}. The example begins with the definition of three concepts
that are used to specify constraints on the type parameters of
\code{foldl}. The \code{Semigroup} concept is used indirectly, as a
nested requirement inside \code{Monoid}. The concept \code{Seq} has
one associated type $E$ for the element type of the sequence. The
\code{foldl} function has one parameter type $S$ for the sequence, and
constrains $S$ to be something that models \code{Seq}. We alias the
element type of the sequence to the name $E$, and constrain $E$ to be
a model of \code{Monoid}. The \code{fix} operator is to make
\code{foldl} a recursive function.  We access the binary operator and
identity element of the monoid and use them to implement the core
logic of the \code{foldl}. Much of the syntactic noise in this part of
the function is because $\mathrm{F}^{\mathcal{G}}$ does not support
function overloading and therefore cannot implicitly introduce concept
members into the scope of a constrained expression without running
into problems with name conflicts. The language $\mathcal{G}$, on the
other hand, adds support for function overloading.

\begin{figure}[tbp]
  \centering
\begin{lstlisting}[morekeywords={in,let,model,fix,type,then}]
concept Semigroup<$\alpha$> { ; ; binary_op : $\alpha->\alpha->\alpha$ } in
concept Monoid<$\alpha$> { ; Semigroup<$\alpha$>; identity_elt : $\alpha$ } in 
concept Seq<$S$> { $E$ ; ; isnull : $S$ $\to$ bool, head : $S$ $\to$ $E$, tail : $S \to S$ } in
let foldl = ($\Lambda$ $S$.  Seq<$S$> $\Rightarrow$ 
           type $E$ = Seq<$S$>.$E$ in 
           Monoid<$E$> $\Rightarrow$
           fix ($\lambda$ r : $S$ $->$ $E$. $\lambda$ ls : $S$.
                let binary_op = Monoid<$E$>.Semigroup<$E$>.binary_op in
                let identity_elt = Monoid<$E$>.identity_elt in
                if Seq<$S$>.isnull ls then identity_elt
                else binary_op(Seq<$S$>.head ls, r(Seq<$S$>.tail ls))))
in
model Semigroup<int> { ; binary_op = $\lambda$ x:int. $\lambda$ y:int. x + y } in
model Monoid<int> { ; identity_elt = 0 } in
model Seq<int list> {E=int; isnull=$\lambda$ ls. null? ls, head=$\lambda$ ls. car ls, tail=$\lambda$ ls. cdr ls} in
foldl[int list] [2,3,4]
\end{lstlisting}
  \caption{Example of a generic fold in $\mathrm{F}^{\mathcal{G}}$.}
  \label{fig:fg-example}
\end{figure}

The dynamic semantics of $\mathrm{F}^{\mathcal{G}}$ is defined through
a type-directed translation to System F, which we refer to here as
$\mathcal{C}(\cdot)$. Without loss of generality, the semantics is
defined for programs (closed expressions) of type $\mathsf{int}$. Let
$\longrightarrow$ be the standard single-step reduction relation for 
call-by-value System F and $\longrightarrow^{*}$ be its reflexive,
transitive closure~\cite{Pierce:2002hj}. Then the dynamic semantics of
$\mathrm{F}^{\mathcal{G}}$ is defined by the following partial function
named $\mathit{eval}$.
\[
  \mathit{eval}(e) = 
  \begin{cases}
    n & \text{if } \mathcal{C}(e) \longrightarrow^{*} n \\
   \uparrow & \text{if } \mathcal{C}(e) \text{ diverges}
  \end{cases}
\]
The translation uses the dictionary-passing approach that is commonly
used for Haskell type classes~\cite{wadler89:_adhoc,hall96:_typeclasses}.
Associated types are translated into extra type parameters.  We do not
give the translation here, but refer the interested reader to the
paper \emph{Essential Language Support for Generic
  Programming}~\cite{Siek:2005mf,siek04:_fg1}.

Getting back to the main point of $\mathrm{F}^{\mathcal{G}}$, does it
provide modular type checking? The answer is yes and that can be seen
by inspecting the type rules in Figure~\ref{fig:fg-type}.  The rules
for functions and generics are the same as those of System F, which we
inspected for modularity in Section~\ref{sec:higher-order-poly}.  The
important addition is the constrained expression $C \Rightarrow e$. The
introduction rule for constrained expressions only refers to the
constraint $C$ and body $e$.  The elimination rule for constrained
expressions only depends on the type of expression $e$, that is, the
type $C \Rightarrow \tau$, and the current environment $\Gamma$.

However, an experienced language designer knows that it is easy to
create a modular type system by cheating, that is, by creating a type
system that does not guarantee type safety.  Thus, the type safety
result for $\mathrm{F}^{\mathcal{G}}$ plays an important role in
verifying that $\mathrm{F}^{\mathcal{G}}$ has a modular type system.
The type safety theorem for $\mathrm{F}^{\mathcal{G}}$ is stated as
follows.

\begin{theorem}[Type Safety]
  If $\vdash e : \mathsf{int}$, then either $\mathit{eval}(e) = n$ for
  some integer $n$ or $\mathit{eval}(e) =\, \uparrow$.
\end{theorem}

System F is type safe, so proving type safety for
$\mathrm{F}^{\mathcal{G}}$ amounts to proving that the translation
$\mathcal{C}$ is type preserving. Indeed, I proved the following
lemma~\cite{Siek:2005mf,siek04:_fg1}.
\begin{lemma}[$\mathcal{C}$ is type preserving]
  \label{lem:type-pres}
  If $\vdash e : \tau$, then $\vdash \mathcal{C}(e) : \mathcal{C}(\tau)$.
\end{lemma}

\noindent The proof of type safety then proceeds as follows.

\begin{proof}[Proof of Type Safety]
  We are given that $\vdash e : \mathsf{int}$.  By
  Lemma~\ref{lem:type-pres}, we have $\vdash \mathcal{C}(e) :
  \mathsf{int}$ (we have $\mathcal{C}(\mathsf{int}) =
  \mathsf{int}$). 
  Then by the type safety of System F,
  $\mathcal{C}(e)$ either reduces to a value of type $\mathsf{int}$,
  in which case the value is an integer, or $\mathcal{C}(e)$
  diverges. In the first case, we have $\mathit{eval}(e) = n$
  and in the second case, $\mathit{eval}(e) =\, \uparrow$. 
  Either way, our proof is complete.
\end{proof}

In addition to having a modular type system,
$\mathrm{F}^{\mathcal{G}}$ supports separate compilation, as
$\mathrm{F}^{\mathcal{G}}$ translates to System F, and it is
straightforward to separately compile System F as we discussed
in Section~\ref{sec:higher-order-poly}.
